\documentclass[aps,twocolumn,superscriptaddress]{revtex4-2}

\usepackage{graphicx}
\usepackage{hyperref}
\usepackage{amsmath,amssymb, mathtools, physics, dsfont}
\usepackage{orcidlink}
\usepackage[export]{adjustbox}
\definecolor{lightgray}{rgb}{0.96,0.96,0.96}
\usepackage{multirow} 

\usepackage{listings}
\lstnewenvironment{python}[1][]{
	\lstset{language=Python,
		basicstyle=\small\ttfamily,
		commentstyle=\small\ttfamily\color{darkgray},
		breaklines=true,
		frame=none,
		backgroundcolor=\color{lightgray},
		upquote=true}
	}
	{}
	
\usepackage[dvipsnames]{xcolor}
\hypersetup{
		colorlinks=true,       
		linkcolor=Brown,        
		citecolor=Blue,        
		filecolor=magenta,     
		urlcolor=RoyalPurple,      
	}

\usepackage{tikz} 
\usetikzlibrary{quantikz}

\begin{document}

\title{Digital Twin Simulations Toolbox of the Nitrogen-Vacancy Center in Diamond}

\author{Lucas Tsunaki \orcidlink{0009-0003-3534-6300}}
\affiliation{Department Spins in Energy Conversion and Quantum Information Science (ASPIN), Helmholtz-Zentrum Berlin für Materialien und Energie GmbH, Hahn-Meitner-Platz 1, 14109 Berlin, Germany}

\author{Anmol Singh \orcidlink{0009-0000-1578-626X}}
\affiliation{Department Spins in Energy Conversion and Quantum Information Science (ASPIN), Helmholtz-Zentrum Berlin für Materialien und Energie GmbH, Hahn-Meitner-Platz 1, 14109 Berlin, Germany}

\author{Sergei Trofimov \orcidlink{0000-0002-9718-2845}}
\affiliation{Department Spins in Energy Conversion and Quantum Information Science (ASPIN), Helmholtz-Zentrum Berlin für Materialien und Energie GmbH, Hahn-Meitner-Platz 1, 14109 Berlin, Germany}

\author{Boris Naydenov \orcidlink{0000-0002-5215-3880}}
\email{boris.naydenov@helmholtz-berlin.de}
\affiliation{Department Spins in Energy Conversion and Quantum Information Science (ASPIN), Helmholtz-Zentrum Berlin für Materialien und Energie GmbH, Hahn-Meitner-Platz 1, 14109 Berlin, Germany}
\affiliation{Department of Physics, Freie Universität Berlin, Arnimallee 14, 14195 Berlin, Germany}

\date{\today}

\begin{abstract}
	
The nitrogen-vacancy (NV) center in diamond is a crucial platform for quantum technologies, where its precise numerical modeling is indispensable for the continued advancement of the field.
We present here a Python library for simulating the NV spin dynamics under general experimental conditions, i.e. a digital twin.
Our library accounts for electromagnetic pulses and other environmental inputs, which are used to solve the system's time evolution, resulting in a physical output in the form of a quantum observable given by fluorescence.
The simulation framework is based on a non-perturbative time-dependent Hamiltonian model, where the states initialization and readout are postulated from the interaction with optical fields.
By eliminating oversimplifications such as the adoption of rotating frames for the microwave and radio frequency fields, our simulations reveal subtle dynamics emerging from realistic pulse constraints.
The software is illustrated with three examples and validated by comparing the simulations with experimental reports, relevant to the fields of quantum computing (conditional logic gates), sensing (dynamical decoupling sequences with coupled spins) and networks (state teleportation).
Overall, this digital twin delivers a robust numerical modeling of the NV spin dynamics, with simple and accessible usability, which can be used for a wide range of applications. 

\end{abstract}

\maketitle


\section{Introduction}

The advancement of the second quantum revolution relies on precise and faithful modeling of quantum components for scalable architectures.
These models must be able to develop and validate the capabilities of the quantum devices in a clear, consistent, and standardized way.
A quantum component can be understood in terms of an element which receives a series of physical inputs and produces a series of outputs, governed by the fundamental laws of quantum mechanics.
Color center defects in solids are a concrete example of such components and thus have been a focal point in quantum technologies research~\cite{CC1, CC2}.

\begin{figure}[b!]
	\includegraphics[width=\columnwidth]{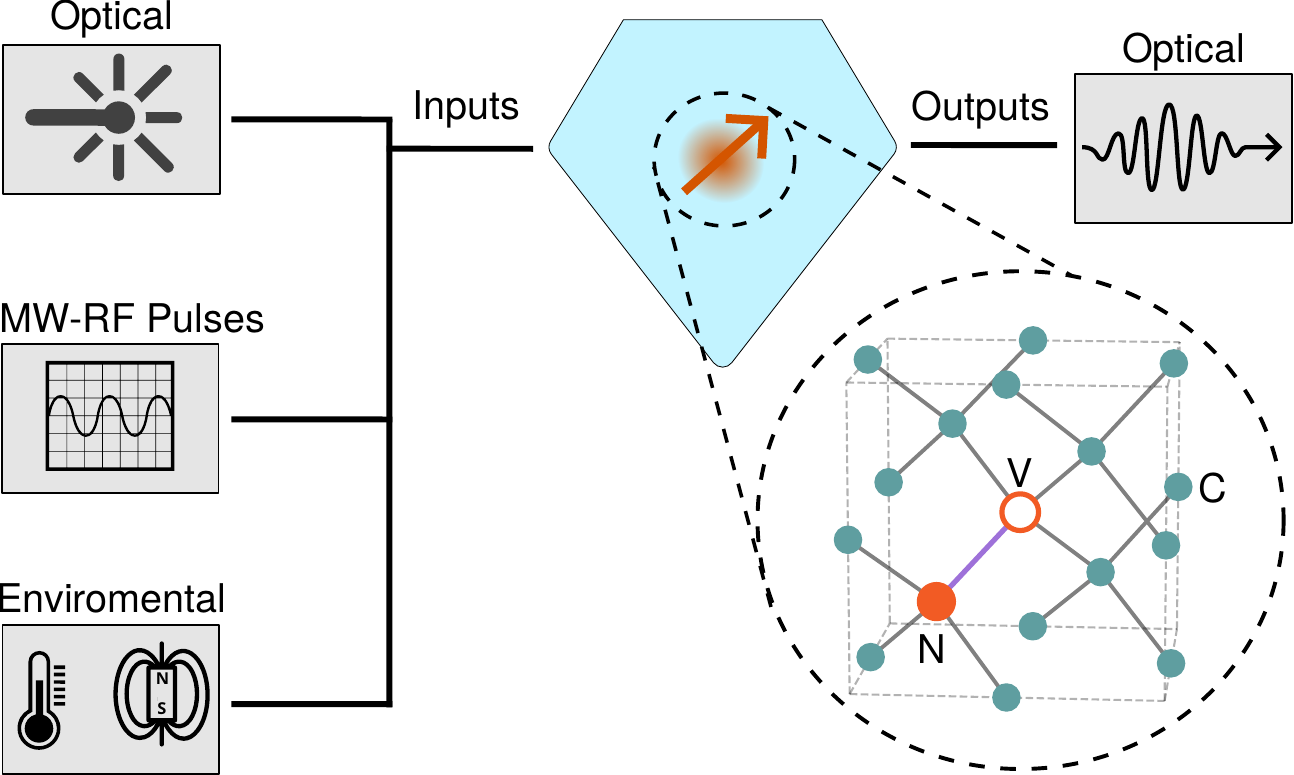}
	\caption{Schematic representation of a color center digital twin.
	Color centers are point defects in solids that strongly interact with light and often possess a quantum-mechanical spin.
	The color center simulations software receives a series of physical inputs and generates an output, which is calculated	numerically based on the time evolution of the system under the system's Hamiltonian.
	For the NV center in diamond, three inputs can be distinguished: optical (for initialization of spins), microwave and radio frequency pulses (for spin control) and environmental (such as temperature and external magnetic field).
	The output of the NV component is fluorescence, which can be related to a quantum-mechanical observable.}
	\label{fig:component_scheme}
\end{figure}

Color centers are impurities in the lattice of solid crystals, which strongly interact with optical fields and often possess magnetic spins described quantum mechanical operators.
This way, three inputs can be defined for a general color center, as represented in Fig.~\ref{fig:component_scheme}.
First, an optical input, which can be either modeled classically, as an oscillating electromagnetic field, or quantum-mechanically, as a photon Fock state. The second input is given by magnetic wave pulses that induce spin level transitions~\cite{rabi}, usually in the microwave (MW) and radio frequency (RF) ranges.
The third input includes interactions with the environment, which influence the dynamics of the system, such as temperature and external magnetic fields. Given these three inputs, quantum mechanics governs the time evolution of the system resulting in a physical output.
This output can take a form of an expectation value of some observable operator, or of a photon emission in the associated Fock space.

Within these considerations, the framework for simulating color centers is rather broad, as each system has its own specific attributes. This study then focuses on the nitrogen-vacancy (NV) center in diamond~\cite{NV_review1, NV_review2}, where the inputs and outputs are constrained to a more concrete simulation framework. Specifically, the NV system enables us to phenomenologically model the interaction with optical fields, by postulating an initial state prepared via an optical pumping mechanism and defining the output as the probability of occupying the bright spin state, which corresponds to the measured spin-dependent luminescence.
The same does not apply to all color centers.
For example, in the silicon vacancy (SiV) in diamond, light can be efficiently used for coherent control of the spin~\cite{SiV_1}. Nonetheless, many aspects of the NV digital twin developed here, such as the interaction with MW and RF pulses, are common to other color centers and quantum systems in general. 

With this phenomenological approach for modeling the interaction of the NV digital twin with optical fields, we avoid simulating the intricate optical processes and significantly simplify the simulation framework by reducing the number of experimental variables within the digital twin.
The latter is already fairly large considering the MW-RF pulses and environmental inputs.
Undoubtedly, this approximative treatment neglects other interesting physical phenomena related to such optical processes of the NV system~\cite{PES, thermo_opt_pumping, optical_pumping, optical_pumping_2}, which thus streamlines a future road map towards a general purpose color center digital twin.

Independent of the system, calculating an output for an arbitrary given input is not always a straightforward task~\cite{QME}, since exact analytical solutions are not usually obtainable.
Perturbative approximations for MW and RF fields, such as the adoptions of rotating frame with rotating wave approximations (RWA)~\cite{RWA_1, RWA_2, slichter}, can be used to some extent to calculate the dynamics of quantum systems semi-analytically.
However, these approximations restrict the operation conditions of the digital twin inputs, by requiring weak driving regimes for the RF-MW pulses input~\cite{strong_driving1, strong_driving_2, beyond_RWA} and internal Hamiltonians which need to commute with the rotating frame operators~\cite{slichter}.
The adoption of rotating frames and RWA can also overlook important physical phenomena, such as phase accumulation of the spins during pulses with finite length~\cite{NV_teleportation_1} and the appearance of ambiguous resonances in multipulse quantum sensing~\cite{ambiguous_resonances}.

Given the three distinct types of inputs, the most appropriate description of the NV digital twin is numerically solving its time evolution under the MW and RF fields in the static laboratory frame, based on quantum master equations~\cite{QME}.
For that, we use Quantum Color Centers Analysis Toolbox (QuaCCAToo)~\cite{quaccatoo}, a Python object-oriented library for modeling the NV, which also allows users to simulate arbitrary quantum systems.
The digital twin is then able to operate with general environmental inputs, with external magnetic fields of arbitrary intensities and orientations, and with temperatures in the range from 5.6~K to 700~K.

We begin this work in Sec.~\ref{sec:principles} by discussing in detail the physical and mathematical modeling of the NV digital twin.
Based on that, in Sec.~\ref{sec:applications}, we provide three applications of the NV simulations software based on previous experimental studies~\cite{polarization_population_2, RWA_2, 13C_sensing_1, ambiguous_resonances, NV_teleportation_1}.
There we showcase and exemplify the use of the software in applications ranging from quantum information processing, to sensing and networks.
We then conclude in Sec.~\ref{sec:conclusion} giving an overview of the NV software component, with its limitations and possible improvements towards a general color center tool.

\section{Physical and Mathematical Principles}\label{sec:principles}

Diamond stands out as an exceptional host material for color centers due to its unique combination of physical and chemical properties.
First, the wide band-gap of 5.47~eV and the possibility to achieve low concentrations of impurities provides unique optical and spin stability for possible defects~\cite{diamond_review}.
Furthermore, the strong carbon bonds of the lattice guarantee unmatched mechanical and chemical stability.
This allows operating in large ranges of pressure (from ultrahigh vacuum up to several GPa~\cite{NV_pressure}) and temperature (from few mK to above room temperatures~\cite{NV_temperature_1, NV_temperature_2}), while being compatible with most chemical and biological systems~\cite{biological_compatibility, cancer1, cancer2}.

\begin{figure*}[t!]
	\includegraphics[width=\textwidth]{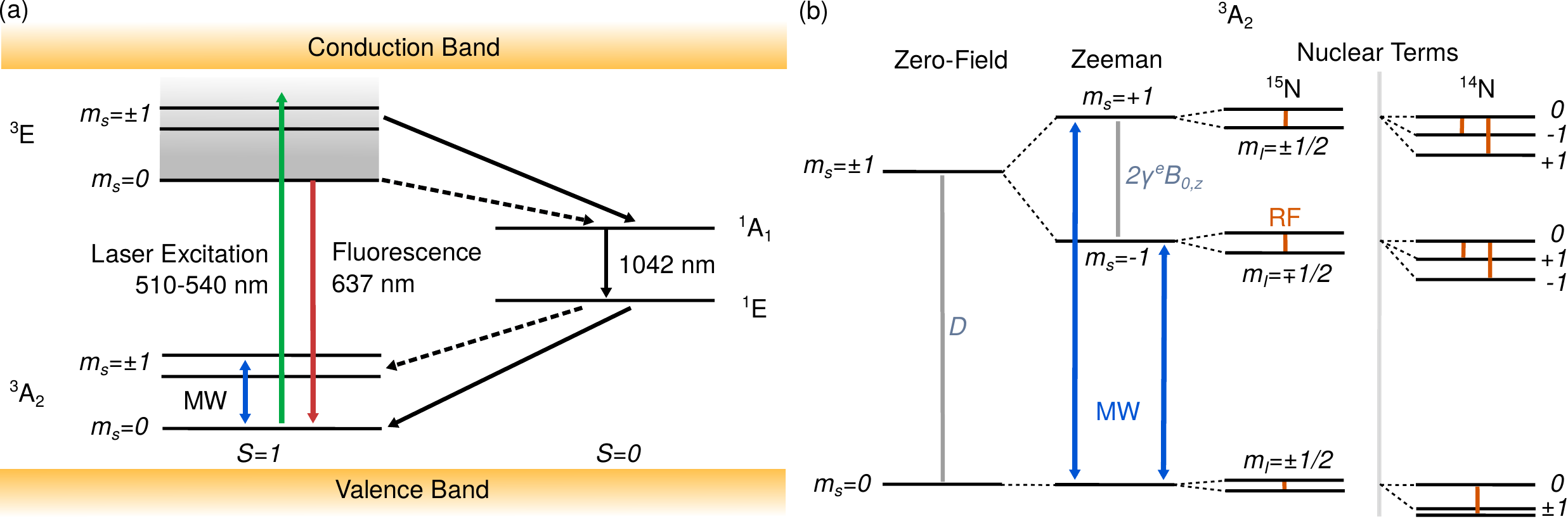}
	\caption{\textbf{(a)} Optical pumping and readout mechanism of NV centers.
    The latter has triplet ($S=1$) and singlet ($S=0$) states within the band gap of diamond. 
    The ground triplet states ($^3$A$_2$) can be driven from thermal equilibrium to the excited triplet states ($^3$E) by a non-resonant green laser.
    From there, the NV can decay back to $^3$A$_2$ through the emission of red fluorescence.
    Alternatively, it can undergo an intersystem crossing through the intermediate singlet states $^1$A$_1$ and $^1$E, leading to polarization of the $m_S=0$ spin sublevel. 
    The same optical pumping mechanism can be used to measure the $m_S$ states populations.
    \textbf{(b)} Energy level diagram of NV$^-$.
    The dominant zero-field term splits the $m_S=0$ and $m_S=\pm1$ states, where the latter are further split by Zeeman interaction in the presence of an external magnetic field $\bold{B}_0$. Resonant MW pulses can drive coherent transitions between $m_S=0 \leftrightarrow \pm1$.
    Depending on the nitrogen isotope, $^{15}$N or $^{14}$N, the levels are further split, where transitions between $m_I$ dependent on $m_S$ can be achieved with resonant RF pulses.}
	\label{fig:energy_levels}
\end{figure*}

Among the various color center defects in diamond~\cite{CC_diamond}, the NV center is the most studied one. It consists of a substitutional nitrogen atom adjacent to a vacancy in the diamond lattice (see Fig.~\ref{fig:component_scheme}).
The NV can exist in different charge states~\cite{NV_charge_state}, but the negatively charged NV$^-$ state, simply referred to as NV here, is the most relevant for quantum technologies.
In this case, the extra valence electron of the nitrogen atom pairs with a lattice electron, forming either a spin triplet ($S = 1$) or singlet ($S = 0$) configuration.
Both configurations give rise to new electronic orbitals with ground and excited states within the band gap of the diamond host~\cite{ab_initio}.
These states are denoted $^3$E and $^3$A$_2$ for the triplet, and $^1$A$_1$ and $^1$E for the singlet, as schematically shown in Fig.~\ref{fig:energy_levels}~(a).
At room temperature, only the sublevels of the ground triplet state $^3$A$_2$ are equally populated and thus actively used as the computational basis.

Given these basic considerations about NV centers, this section is structured as follows.
In Sec.~\ref{sec:H0}, we introduce the Hilbert space representation of the system and its internal time-independent Hamiltonian $\hat{H}_0$.
In Sec.~\ref{sec:optical}, the optical initialization and readout are discussed and modeled, postulating an initial state and a quantum mechanical observable.
In Secs.~\ref{sec:H1} and \ref{sec:H2}, we present the Hamiltonians describing the interactions with the external control field $\hat{H}_1(t)$ and sensing of spins or oscillating fields $\hat{H}_2(t)$, respectively.
Finally, in Sec.~\ref{sec:dynamics}, we discuss the time evolution dynamics of the system using the Lindblad master equation~\cite{QME} and the numerical methods for solving the resulting differential equations.

\subsection{Internal Hamiltonian $\hat{H}_0$}\label{sec:H0}

The NV center has an intrinsic nuclear spin from the nitrogen nucleus, which can be either $I^n=1/2$ for the $^{15}$N isotope or $I^n=1$ for $^{14}$N, apart from the electron spin.
Regardless, the NV state is represented in the composite Hilbert space of the two spins: $\mathcal{H}_{NV} = \mathcal{H}_{S} \otimes \mathcal{H}_{I^n}$.
The center can further couple with other spins, as discussed in Sec.~\ref{sec:H2}, which adds another sub-space to the total Hilbert space.
For other color centers or in applications where the optical field is quantized, both not treated here, the total space is further composed with the photon Fock space of continuous wavelengths $\lambda$: $\mathcal{H}_{NV} \otimes \mathcal{F}$.

The ground triplet state $^3$A$_2$ can be modeled by a Hamiltonian~\cite{NV_hamiltonian1, NV_hamiltonian2, NV_hamiltonian3} using the electron and nitrogen nuclear spins operators $\hat{S}$ and $\hat{I}$ in terms of decreasing energy as
\begin{multline}\label{eq:H0}
	\hat{H}_{0} = D \hat{S}_z^2
	- \gamma^e \bold{B}_0 \cdot \hat{\bold{S}}
	+ a_\parallel^n \hat{S}_z \hat{I}_z^n
	+ a_\perp^n \left( \hat{S}_x \hat{I}_x^n + \hat{S}_y \hat{I}_y^n \right) \\
	-  \gamma^n \bold{B}_0 \cdot \hat{\bold{I}}^n + Q (\hat{I}_z^n) ^2.
\end{multline}
Here, the $z$-axis is taken along the NV symmetry axis and $\bold{B}_0$ is the external static magnetic field, being part of the environmental inputs.
We also consider the Hamiltonians in units of frequency through the text, or in other words, the Planck constant is $h = 1$.
The first and also the dominant term at $|\bold{B}_0|<102.4$~mT originates from the dipolar spin-spin interaction between the two unpaired electrons, resulting in a zero-field splitting between $m_S=0$ and $m_S=\pm1$ of $D=2.87$~GHz at room temperature.
The value of $D$ changes slightly with temperature, as the dipolar interaction between the two electron spins depends on the lattice constant, which is temperature dependent~\cite{ms_T_dependence_1, ms_T_dependence_2, ms_T_dependence_3}.
To model the temperature dependence of $D$, we use two different phenomenological polynomial equations.
The first one being of 5\textsuperscript{th} order with the temperature in range the from 5.6~K to 295~K, as measured by X. D. Chen \textit{et al.} (2011)~\cite{ms_T_dependence_2}, and the other one of 3\textsuperscript{rd} order in the temperature range of 295~K to 700~K, as measured by D. M. Toyli \textit{et al.} (2012)~\cite{ms_T_dependence_3}.
The maximum temperature of operation is mostly limited by the optical initialization and readout mechanism, as will be discussed in the following section.

The second term corresponds to the Zeeman interaction of the electron spin with the external field ($\gamma^e=-28.025$~GHz/T), further splitting the $m_S=\pm1$ levels.
At intense magnetic fields above $B_\text{GSLAC} = 102.4$~mT, the ground state level anticrossing (GSLAC)~\cite{LAC, LAC_2}, the Zeeman term dominates over the zero-field splitting and the $m_S=-1$ state becomes less energetic than $m_S=0$.
For magnetic fields that are not aligned with the NV axis, this term also causes state mixing of the electronic states~\cite{state_mixing}.
The third and fourth terms represent the parallel and perpendicular components of the hyperfine interaction between the two spins, with $a_\parallel^n=-2.14$~MHz and $a_\perp^n=-2.70$~MHz for $^{14}$N, $a_\parallel^n=3.03$~MHz and $a_\perp^n=3.65$~MHz for $^{15}$N~\cite{Newton09}.
Analogous to the electron spin, the fifth term describes the Zeeman interaction of the nuclear spin with $\bold{B}_0$, with the gyromagnetic ratios of $\gamma^n = -4.316$~MHz/T for $^{15}$N and $\gamma^n = 3.077$~MHz/T for $^{14}$N.
Lastly, the last term represents the quadrupole interaction of the nuclear spin ($Q=-5.01$~MHz), only present in the $^{14}$N isotope. The complete energy level structure of the $^3$A$_2$ state for both nitrogen isotopes at magnetic fields smaller than $B_\text{GSLAC}$ is shown in Fig.~\ref{fig:energy_levels}~(b).

Additional terms are also present in the ground state Hamiltonian when considering the effects from external electric fields and lattice strain~\cite{NV_review1, 3A_theory}.
The former interaction with the NV is particularly useful for electric field sensing~\cite{E_sensing}, while the lattice strain terms are relevant in nanodiamonds, which are essential for nanoscale quantum sensing applications~\cite{biological_compatibility, ms_T_dependence_1}.
Although these terms are not yet implemented in QuaCCAToo yet, they can be easily added to the predefined Hamiltonian by the user. 
 
In many cases (Sec.~\ref{sec:rotations}), the nitrogen nuclear spin terms can be neglected due to their magnitudes being much smaller than the electron spin ones.
However, these terms are a fundamental part of the NV, serving as a basis for other applications (Secs.~\ref{sec:sensing} and ~\ref{sec:network}).
Depending on the goal of the application, it may be preferable to have the $^{15}$N or $^{14}$N isotope~\cite{ambiguous_resonances}.
The $^{14}$N isotope has a much larger natural abundance (99.6\%) when compared to $^{15}$N (0.4\%), but either one can be precisely selected in purified chemical vapor deposition~\cite{CVD, delta_doping} or ion implantation~\cite{implantation, delta_doping} fabrication methods.


\subsection{Optical Initialization and Readout}\label{sec:optical}

A requirement for quantum computing and other quantum technology applications is the ability to reliably initialize and readout quantum states~\cite{divincenzo}.
An important feature of NVs is the presence of a simple optical pumping mechanism~\cite{optical_pumping} capable of initializing the electronic spin, in addition to providing a viable measurement method, as represented in Fig.~\ref{fig:energy_levels}~(a).
By applying a non-resonant green laser of wavelengths ranging from 510 to 540~nm (green solid arrow), the ground triplet states $^3$A$_2$ can be driven to the excited triplet states $^3$E in a spin-conserving transition.
From there, the NV can rapidly disperse energy to the diamond lattice, decaying back to the ground triplet via the emission of red fluorescence (red solid arrow), while preserving its spin state $m_S$.
This results in a broad photoluminescence spectrum with a characteristic zero-phonon line at 637~nm~\cite{NV_PL}.

Alternatively, the excited NV can undergo a non-spin-conserving decay with a photon emission of 1042~nm through the singlet states, known as intersystem crossing (ISC).
The excited states with $m_S=\pm1$ (black solid arrow) are more likely to decay through this path than $m_S=0$ (black dashed arrow). Which in turn are more probable to decay back to $m_S=0$ (black solid arrow).
This way, after a few \textmu{s}, the electron spin is polarized in a pseudo-pure state due to the limited polarization efficiency, with a normalized population in the $m_S=0$ level of $n_0$ typically greater than  0.6~\cite{rabi,polarization_population_2} and up to 0.9~\cite{RWA_2}.
This can be further increased to values above $n_0>0.98$ using the level anticrossing~\cite{optical_pumping_2}.
The exact polarization of the populations can also be calculated from the rate equations of the transitions~\cite{polarization_population}.

The nitrogen nuclear spin, in contrast to the electron spin, is typically in a thermal equilibrium mixed state given by its Boltzmann distribution.
This way, the initial density matrix state of the $^{15}$NV system in the Hilbert space representation introduced in Sec.~\ref{sec:H0}, for a given population $n_0$, is written as
\begin{equation} \label{eq:BD_rho0}
	\hat{\rho}_0 = \frac{1}{2} \begin{bmatrix}
		1 - n_0 & 0 & 0 \\
		0 & 2n_0 & 0 \\
		0 & 0 & 1 - n_0
	\end{bmatrix} 
	\otimes \frac{1}{Z}
	\begin{bmatrix}
		e^{-\beta E_0} & 0 \\
		0 & e^{-\beta E_1}
	\end{bmatrix} ,
\end{equation}
where $Z = e^{-\beta E_0} + e^{-\beta E_1}$ is the partition function, $\beta = 1/k_B T$, and $k_B$ is the Boltzmann constant.
Here the temperature $T$ appears as another environmental input for the digital twin.
$E_0$ and $E_1$ are the energy eigenvalues of the nitrogen nuclear spin at $m_S=0$ calculated from the Hamiltonian in Eq.~\ref{eq:H0}.
At room temperature and moderate fields, the thermal energy is much larger than the states-separation, $|E_1 - E_0| \ll k_B T$, and the density matrix can be well-approximated by the identity matrix.
Furthermore, an initial $m_S=0$ population smaller than 1 affects mostly the resulting contrast of the the observable obtained after a protocol, while it slows down numerical simulations by involving density matrices with more elements (Sec.~\ref{sec:dynamics}).
In most cases then, the initial state can be well approximated by
\begin{equation}\label{eq:rho0}
	\hat{\rho}_0 \cong \ket{0}\bra{0} \otimes \frac{\hat{\mathds{1}}}{2},
\end{equation}
where $\ket{0} \equiv \ket{m_S=0}$.
For the $^{14}$NV center, the same considerations are still valid, but with a higher dimension density matrix.
In applications where initialization of the nuclear spin is required, the polarization of the electron spin can be transferred to it through SWAP-like operations~\cite{N_gates} and dynamical nuclear polarization~\cite{DNP}.

Apart from enabling initialization of the electron spin, the same optical pumping mechanism can be used for reading out the $m_S$ populations.
As the $m_S=\pm1$ states decay more likely through the singlet states than $m_S=0$, an increase in their populations results in a decrease of the fluorescence intensity.
This way, we define the fluorescence observable operator simply as
\begin{equation}\label{eq:F_S}
	\hat{F}_S = \ket{0}\bra{0} \otimes \hat{\mathds{1}}.
\end{equation}
This observable is related to the $\hat{S}_z$ operator, but not strictly the same, because the fluorescence operator cannot be used to determine the difference between the $m_S=\pm1$ states.
The exact relation between the experimentally measured fluorescence and the expectation value of the observable depends on the fundamental probability for the ISC~\cite{PES, thermo_opt_pumping} and several other experimental parameters, such as the photon collection efficiency, background counts, laser power and readout time.
Rather than simulating all of these effects, we choose a normalized observable and then we post-process the experimental data with the \texttt{ExpData} class from QuaCCAToo, which is compatible with the data structure from the Qudi experimental control software~\cite{qudi}.

The NV does not present a reliable direct readout mechanism for the nuclear spin.
However, the latter can be measured with a single shot readout technique through repetitive conditional excitations of the electron spin~\cite{single_shot_readout}.
In addition, it is important to note that this optical initialization and readout mechanism is limited by the non-radiative processes experienced by the NV~\cite{ms_T_dependence_3}, which restricts the temperature range of operation of the digital twin to values below 700~K.

In this work, the optical input and output of the NV digital twin are not actually simulated, but instead impose postulates regarding the states initialization and measurement observable.
Therefore, the photon emission spectrum properties of the NV, such as the temperature dependent phonon mediated processes~\cite{PE_spectrum_1, PE_spectrum_2}, are not reproduced within this model.
Such photon emission statistics of NVs can be simulated using other models~\cite{PES} that focus on the interaction of the NV center with optical fields based on first principle calculations~\cite{NV_optical_interaction}. This interaction of the quantum mechanical spin with light, characteristic to color centers, enables us to efficiently polarize and readout the spin state of single NV centers. A remarkable improvement when compared to conventional electron paramagnetic and nuclear magnetic resonances (NMR), which require ensembles of spins for precise detection and usually rely on thermal polarization.


\subsection{Control Field Hamiltonian $\hat{H}_1(t)$}\label{sec:H1}

Another requirement of quantum technologies is the ability to realize a universal set of quantum gates between the qubit states~\cite{divincenzo}.
For some color centers with small spectral diffusion, as the SiV~\cite{SiV_1}, this can be easily achieved with light.
For NVs however, the energy difference between $m_S$ states of the ground state triplet $^3$A$_2$ is in the MW frequency range.
MW excitation is therefore the most straightforward method for having coherent control of the electron spin.
Similarly, transitions of the nitrogen nuclear spin are achieved with RF pulses.

For a general time-dependent excitation field $\bold{B}_1(t)$, the control Hamiltonian in the $\mathcal{H}_{NV}$ space is given by
\begin{equation}\label{eq:H1}
	\hat{H}_1(t) = - \bold{B}_1(t) \cdot \left( \gamma^e \bold{\hat{S}} + \gamma^n \bold{\hat{I}}^n \right)
\end{equation}
This definition allows our software to operate with general excitation fields, such as with arbitrary polarizations~\cite{strong_driving_2} or non-harmonic and non-square time dependencies~\cite{optimal_control_1, optimal_control_2}, as calculated using optimal control theory.
When the NV is coupled with an additional spin, as discussed in the following section, another operator term is added for the corresponding spin.

While Eq.~\ref{eq:H1} describes a general framework for the NV interaction with external control fields, most applications assume linearly-polarized square harmonic pulses in the form of
\begin{equation}\label{eq:B1}
	\bold{B}_1(t) =  \sum_k B_{1,k} \cos(\omega_{0,k} t + \phi_k) \; \text{rect} \left(\frac{t - t_{0,k}}{t_{p,k}}\right) \bold{x}.
\end{equation}
Here, the field is assumed to be along the $\bold{x}$ direction, perpendicular to the quantization axis $z$.
This represents a sum of harmonic pulses, each with frequency $\omega_{0,k}$, intensity $ B_{1,k}$ and phase $\phi_k$.
Each harmonic function is then modulated by a rectangular pulse envelope defined as
\begin{equation*}
	\text{rect}(t)  = 
	\begin{cases} 
		1 & \text{if } 0 \leq t < 1, \\
		0 & \text{otherwise},
	\end{cases}
\end{equation*}
with $t_{p,k}$ being the pulse length and $t_{0,k}$ its initial position in time.
Under such harmonic pulses, Eq.~\ref{eq:H1} reduces to
\begin{equation*}
	\hat{H}_1(t) = - \sum_k \cos(\omega_{0,k} t + \phi_k) \; \text{rect} \left(\frac{t - t_{0,k}}{t_{p,k}}\right) \hat{h}_{1,k}.
\end{equation*}
For convenience, we also define the time-independent operator
\begin{equation}\label{eq:h1}
    \hat{h}_{1,k} \equiv \omega_{1,k}^e \sqrt{2}\hat{S}_x + \omega_{1,k}^n 2\hat{I}_x ,
\end{equation}
with the so called Rabi frequencies~\cite{rabi} $\omega_{1,k}^e=\gamma^e B_{1,k}/\sqrt{2}$ and $\omega_{1,k}^n=\gamma^n B_{1,k}/2$.
The $\sqrt{2}$ and 2 normalization factors in each Rabi frequency come from the relation between the spin operators $\bold{\hat{S}}$ and $\bold{\hat{I}}$ with the Pauli matrices $\hat{\sigma}$ for dimensions 3 and 2 respectively.

By taking the pulse frequency of the excitation field $\omega_{0,k}$ in resonance with one the electron transition [blue arrows in Fig.~\ref{fig:energy_levels}~(b)] or the nuclear transitions (orange lines), rotations of the corresponding spin can be achieved.
There, the phase $\phi_k$ controls the rotation axis in the rotating frame~\cite{slichter} and the pulse duration $t_{p,k}$ controls the rotation angle $\theta_k = 2\pi \omega_{1,k} t_{p,k}$, with $\omega_{1,k}$ hence being the corresponding frequency for the transition, i.e. the Rabi frequency.
Finally, by changing the pulse parameters, a complete set of unitary rotation operators $\hat{R}_\phi(\theta)$ with near unity fidelities can be achieved for both spins of the NV~\cite{N_gates}.

In most experimental conditions, we can assume that the RF field does not interact with $\hat{S}$ and the MW does not interact with $\hat{I}$, due to the large difference in their frequencies -- an approximation which substantially increases computational performance.
However, for magnetic fields around the GSLAC, the resonant frequencies for the electron spins are on the same order than the nuclear spins and this approximation is no longer valid.
Both regimes can be simulated with the digital twin.
In Sec.~\ref{sec:rotations}, we show an application without this approximation (even if the magnetic field is not close to GSLAC) and in Secs.~\ref{sec:rotations} and \ref{sec:network}, we make use of the approximation for the improved computational performance.

Typically, rotating frames are adopted~\cite{slichter} to remove the time-dependency of Eq.~\ref{eq:B1}.
Nevertheless, the internal Hamiltonian $\hat{H}_0$ (Eq.~\ref{eq:H0}) does not commute with the rotation operators and even though the non-secular terms can be treated as a perturbation to the dominant terms~\cite{13C_1, RWA_1, RWA_2, RWA_3}, this approach overlooks important effects caused by the non-commuting terms~\cite{ambiguous_resonances}.
The adoption of the rotating frame and RWA also restricts the range of operation of the MW-RF pulses software input~\cite{double_quantum, strong_driving_2}.
The most general description of the NV interaction with the control field is therefore given in the static laboratory frame, where the precession of each spin with different Larmor frequencies need to be accounted for, as treated in Sec.~\ref{sec:network}.


\subsection{Sensing Interaction Hamiltonian $\hat{H}_2(t)$}\label{sec:H2}

One of the main applications of NV centers is in quantum sensing~\cite{quantum_sensing}.
Differently from classical sensing, a quantum probe is inseparable from the sensed system, where the sensor is affecting and being affected by its dynamics.
Consequently, the sensed system must be incorporated into the Hamiltonian model of the NV.

First, let us consider that the NV is coupled to a spin of arbitrary dimension, denoted as $I^c$.
This spin could be related, for instance, to a $^{13}$C nuclear spin in the diamond lattice, an external spin on the surface, or some paramagnetic defect as P1 centers~\cite{DEER_NV}.
Independent of the origin of this spin, the interaction with the electron spin will be much stronger than with the nitrogen nuclear spin, given the disparity in the gyromagnetic ratios (Sec.~\ref{sec:H0}).
Hence, the interaction between the target and the nitrogen nuclear spins can be neglected, resulting in a Hamiltonian written in the $\mathcal{H}_{NV} \otimes \mathcal{H}_{I^c}$ space as
\begin{equation}\label{eq:H2_quantum}
	\hat{H}_2 = \bold{\hat{S}} \cdot \bold{A}_{hf}^c \cdot \bold{\hat{I}}^c + \hat{H}^c_0 .
\end{equation}
The coupling between the electron spin operator and the sensed spin is described in terms of the hyperfine coupling tensor $\bold{A}_{hf}^c$, which needs to be symmetric.
$\hat{H}^c_0$ describes the internal time-independent Hamiltonian of the sensed spin.
Depending on the structure of the coupling, the energy levels shown in Fig.~\ref{fig:energy_levels}~(b) are further split.
This representation can be used to describe both single spins coupled to the NV, or an effective approximate coupling with ensembles~\cite{ambiguous_resonances}.

Apart from spins, the NV can also be used to sense weak oscillating fields aligned with its quantization axis $\bold{z}$.
In this case, we consider a classical harmonic field with frequency $\omega_2$ given by $\bold{B}_2(t) = B_2 \cos(\omega_2 t)\, \bold{z}$ as the sensed quantity.
The interaction is thus described semi-classically in the $\mathcal{H}_{NV}$ space as
\begin{equation}\label{eq:H2_classical}
	\hat{H}_2(t) = - \gamma^e B_2 \cos(\omega_2 t)\, \hat{S}_z ,
\end{equation} 
analogous to Eq.~\ref{eq:H1}.
Even though Eqs.~\ref{eq:H2_quantum} and~\ref{eq:H2_classical} describe different physical processes, an analogy can be drawn between them.
The spin $I^c$ precesses around $\bold{B}_0$, generating an oscillating classical field which can be approximately modeled by $\bold{B}_2(t)$.
On the other hand, a coupled classical oscillating field can be
approximated by an effective hyperfine coupling~\cite{ambiguous_resonances}.
This way, both approaches can be used interchangeably within a certain approximation degree, depending on whether the interaction is better described in terms of the $\bold{A}_{hf}^c$ and $\hat{H}^c_0$ parameters, or using the external magnetic field $\bold{B}_2(t)$.
In the quantum case, the dynamics needs to be computed in an expanded Hilbert space.
Meanwhile, in the semi-classical picture, the increased computational cost arises from the time dependency of the calculation also during free evolutions of the system, which would otherwise be obtained by a single matrix multiplication of the time evolution operator (Sec.~\ref{sec:dynamics}).


\subsection{State Evolution Dynamics}\label{sec:dynamics}

A complete time evolution description of the NV software component given the inputs and outputs must account for the time dependencies of $\hat{H}_1(t)$ (Eq.~\ref{eq:H1}) and $\hat{H}_2(t)$ (Eq.~\ref{eq:H2_classical}) in the laboratory frame.
The dynamics must also be applicable to mixed states described by the density matrix $\hat{\rho}(t)$.
Altogether then, the time evolution of the system can be modeled by the Liouville-von Neumann equation:
\begin{equation}\label{eq:LvN}
	\frac{\partial \hat{\rho}(t)}{\partial t} = - i \left[\hat{H}_0 + \hat{H}_1(t) + \hat{H}_2(t), \hat{\rho}(t) \right] .
\end{equation}
This gives a set of $d^2$ coupled ordinary differential equations (ODEs) for the elements of $\hat{\rho}(t)$, where $d$ is the dimension of the total Hilbert space.
This number is reduced, however, by considering that $\hat{\rho}(t)$ must have a trace equal to 1 and be Hermitian, that is $\hat{\rho}(t)=\hat{\rho}^\dagger(t)$.
Thus, obtaining the time evolution of the system is reduced to a mathematical problem of numerically solving a system of coupled ODEs for each pulse and free evolution between pulses.
For this purpose, we use linear-multistep Adams-type numerical methods~\cite{adams_1, adams_2, adams_3} for solving the ODEs in discretized small time steps.
The implementation of this in QuaCCAToo is built on top of the master equation solver provided by Quantum Toolbox in Python (QuTip)~\cite{qutip_1, qutip_2}.
Which in turn uses the optimized and BLAS (Basic Linear Algebra Subroutines) accelerated matrix algebra subroutines provided by SciPy and NumPy~\cite{scipy}.
This way, the NV digital twin offers accessibility through a high level abstraction and human-readable syntax of Python, however this comes at the cost of poorer efficiency compared to compiled programming languages such as C++ or Fortran.
This performance issue can be mitigated up to some extent by solving the time evolution of the system under different variables (such as pulse length, free evolution times or pulse frequency) at the same time in parallel over multiple CPU cores.
However, one is still bounded by the single threaded nature of ODEs, since the state ahead in time depends on the previous state.

Finally, for a simulated final state $\hat{\rho}(t_f)$, the quantum mechanical observable related to the fluorescence at the end of the pulse sequence is simply given by
\begin{equation}\label{eq:F}
	\langle \hat{F}_S \rangle  = \Tr[ \hat{F}_S \hat{\rho}(t_f) ] ,
\end{equation}
with the fluorescence observable as defined in Eq.~\ref{eq:F_S}.
This expectation value is also denominated as the transition probability for the state $m_S=0$.

In the Liouville-von Neumann equation, non-unitary effects of the system's interaction with its environment are not accounted for.
For that, the time evolution can be generalized with the Lindblad equation~\cite{lindblad} by introducing a dissipative term in the right-hand side of Eq.~\ref{eq:LvN} as
\begin{equation}\label{eq:lindblad}
	\sum_k \Gamma_k \left[ \hat{C}_k \hat{\rho} (t) \hat{C}_k^\dagger - \frac{1}{2} \left\{\hat{\rho} (t),  \hat{C}_k^\dagger \hat{C}_k\right\} \right],
\end{equation}
valid for Markovian stochastic environments~\cite{markov_env}.
Each source of the non-unitary processes is described in terms of collapse operators $\hat{C}_k$ and their respective rates $\Gamma_k$.
For instance, taking $\hat{C}_\pm = \hat{S}_\pm$ with rates $\Gamma_\pm = 1 / \sqrt{2T_1}$ leads to the process of longitudinal relaxation of
the spin sublevels of the NV.
In addition, $\hat{C}_z = \hat{S}_z$ with $\Gamma_z = 1 / \sqrt{T_2}$ corresponds to transverse relaxation or decoherence.

\section{Simulation Applications}\label{sec:applications}

Having introduced the foundational mathematical and physical principles of the NV digital twin, we now exemplify the software usage and expand the physical discussion of its results through three applications based on previous well-established experimental work. First, in Sec.~\ref{sec:rotations}, we introduce a simple example of conditional rotations between the electron spin of the NV and a strongly coupled spin from a $^{13}$C nuclei~\cite{polarization_population_2}.
In succession, in Sec.~\ref{sec:sensing}, we provide a more elaborate example, where Hahn echo~\cite{hahn_echo}, Carr-Purcell-Meiboom-Gill (CPMG)~\cite{CP_sequence, CPMG, DD_NV} and XY8~\cite{XY8} sequences are used to detect and control external spins~\cite{13C_sensing_1, RWA_2}, relevant for quantum computing and sensing applications. 
Lastly, in Sec.~\ref{sec:network}, we explore a quantum teleportation protocol between two NV centers~\cite{NV_teleportation_1}, a foundation for quantum communication protocols.


\subsection{Two-Qubit Conditional Quantum Gates}\label{sec:rotations}

Conditional gates form the building block for quantum information processing applications involving more than one qubit. 
In the NV's case, the energy level structure allows to realize them straightforwardly through selective MW and RF pulses [Fig.~\ref{fig:energy_levels}~(b)].
To begin illustrating the use of the NV software component, we introduce a simple application of a two-qubit conditional gate between the electron spin of the NV and a nuclear spin from a strongly coupled $^{13}$C from the diamond lattice, first observed by F. Jelezko \textit{et al.} (2004)~\cite{polarization_population_2}.

First, the NV system is represented by an instance of the \texttt{NV} class for a given external magnetic field $\bold{B}_0$.
This can be obtained simply through the following code snippet:
\begin{python}
# System definition
from quaccatoo import NV

sys = NV(B0=200, units_B0='mT', N=0)
\end{python}
where the parameter \texttt{N} representing the nitrogen isotope is set to 0.
This way, we neglect the nitrogen nuclear spin, which is not used in this application.
The external magnetic field is assumed to be aligned with the NV axis and to have a magnitude of $B_0 = 200$~mT.
The \texttt{NV} class internally calculates the system's observable $\hat{F}_S$ (Sec.~\ref{sec:optical}) and Hamiltonian $\hat{H}_0$ (Sec.~\ref{sec:H0}), in units of MHz by default.
This way, the time units are in \textmu{s}.

So far, only the electron spin of the NV is defined.
To add the $^{13}$C spin to the system, the $\hat{H}_2$ Hamiltonian needs to be defined as in Eq.~\ref{eq:H2_quantum}.
In this case, the internal Hamiltonian of the $^{13}$C spin is simply given by a Zeeman interaction along the $z$-axis, while the hyperfine coupling between the two spins can be expressed in terms of a single component of $a_{zz}^c=130$~MHz~\cite{polarization_population_2}, even though other spatial components might exist.
Hence,
\begin{equation*}
	\hat{H}_2 = a_{zz}^c \hat{S}_z \hat{I}^c_z - \gamma^c B_0 \hat{I}^c_z
\end{equation*}
with gyromagnetic ratio $\gamma^c= 10.7084\times 10^{-3}$~MHz/mT.
To model this interaction and to add the carbon spin to the NV, the \texttt{add\_spin} method can be used as follows:
\begin{python}
# H2 Hamiltonian and initial state definition
GAMMA_C = 10.7084e-3
azz = 130

H2 = azz*tensor(jmat(1,'z'),jmat(1/2,'z')) - GAMMA_C*sys.B0*tensor(qeye(3),jmat(1/2,'z'))

sys.add_spin(H2)

sys.rho0 = tensor(basis(3,1),basis(2,0))
\end{python}
with \texttt{tensor} representing the tensor product, \texttt{jmat} giving the spin matrices, \texttt{basis} the $Z$ basis states, and \texttt{qeye} the identity matrices, all imported from QuTiP.

In contrast to Eq.~\ref{eq:rho0}, the initial state of the nuclear spin in this application is polarized, where $\ket{m_S=0}\otimes\ket{m_{I^c}=+1/2}$ is set to the \texttt{rho0} attribute as shown above, following QuTiP's basis convention with \texttt{basis(3,1)} representing $\ket{m_S=0}$ and \texttt{basis(2,0)} representing $\ket{m_{I^c}=+1/2}$.
Therefore, the states are pure and can be represented as kets, instead of density matrices.
This makes the simulation computationally less costly, due to the fewer number of ODEs to be solved in the ket state when compared to the whole density matrix.

\begin{figure}[b!]
	\includegraphics[width=0.69\columnwidth, valign=m]{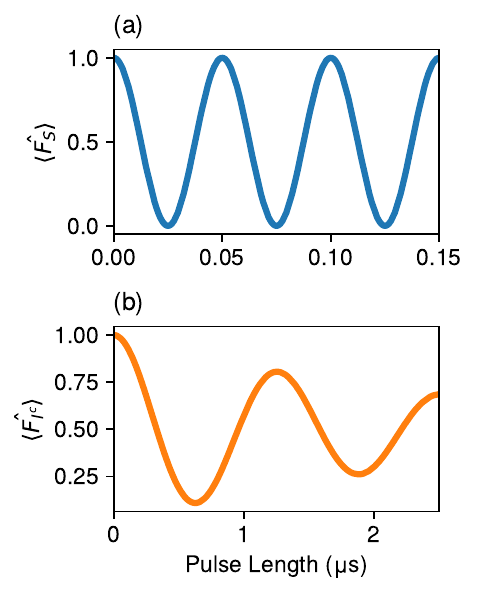}
	\includegraphics[width=.295\columnwidth, valign=m]{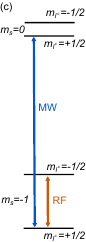}
	\caption{Simulated conditional rotations on \textbf{(a)} NV's electron spin $S$ and \textbf{(b)} nuclear spin $I^c$ from coupled $^{13}$C.
	\textbf{(c)} Energy level diagram of the system, neglecting the $m_S=+1$ state.
	By applying a resonant pulse with either one of the transitions, full Rabi oscillations are achieved to each spin, conditioned to the other.
	The decoherence of the nuclear spin is modeled by a collapse operator within the Lindblad master equation.
	The simulations reproduce the experimental results from F. Jelezko \textit{et al.} (2004)~\cite{polarization_population_2}.}
	\label{fig:rabi}
\end{figure}

The energy levels of the system can be plotted with the \texttt{plot\_energy} method, a sample output of that is shown in Fig.~\ref{fig:rabi}~(c).
At such magnetic field above GSLAC $B_0 > B_{\textrm{GSLAC}}$ (Sec.~\ref{sec:H0}), the $m_S=0$ state has a higher energy than $m_S=-1$, in contrast to the situation depicted in Fig.~\ref{fig:energy_levels}~(b).
Furthermore, the nuclear spin states at $m_S=0$ are only split by their Zeeman interaction, while the sublevels at $m_S=-1$ have a contribution from the hyperfine coupling as well.
In this application, it is important that the $\ket{m_S=0} \leftrightarrow \ket{m_S=-1}$ transitions have very distinct frequencies for $m_{I^c}=+1/2$ and $m_{I^c}=-1/2$ nuclear states, in order to avoid unwanted excitations.
This is ensured by the high intensity magnetic field $B_0$ and hyperfine coupling $a^c_{zz}$, which leads to these energy differences being much larger than the bandwidth of the pulses.

The interaction with an external excitation field needs to be represented through the operator $\hat{h}_1$, as in Eq.~\ref{eq:h1}.
Assuming the Rabi frequencies of $\omega_{1,S} = 20$~MHz and $\omega_{1,I} = 0.8$~MHz for the electron and nuclear spins, respectively (as in Ref.~\cite{polarization_population_2}), we have:
\begin{python}
# h1 Hamiltonian definition
w1_S = 20
w1_I = 0.8	

h1 = w1_S*tensor(jmat(1,'x')*2**.5,qeye(2)) + w1_I*tensor(qeye(3),jmat(1/2,'x')*2)

w0_mw = sys.energy_levels[2]
w0_rf = sys.energy_levels[1]
\end{python}
where we also define the frequencies of the pulses \texttt{w0\_mw} and \texttt{w0\_rf} in resonance with the transitions as shown in Fig.~\ref{fig:rabi}~(c).
By default, in QuaCCAToo, the lowest energy state in the system is 0, that is \texttt{sys.energy\_levels[0]=0}.
In this example, we take the same Hamiltonian $\hat{h}_1$ for both MW and RF fields (Sec.~\ref{sec:H1}).
However, in the subsequent examples, we assume different control Hamiltonians for the two spins for improved computational performance, given their large frequency difference.

Finally, for simulating this interaction of the excitation with the electron spin, we use the \texttt{Rabi} class and the \texttt{square\_pulse} function to model the $\bold{B}_1(t)$ field.
Typically, in a Rabi experiment, the MW (or RF) pulse length is varied, which causes a periodic oscillation of the observable with frequency $\omega_1$.
Given the necessary parameters defined previously and an array for the pulse durations, this is achieved by:
\begin{python}
# Rabi electron simulation
from quaccatoo import Rabi, square_pulse

tp_S = np.linspace(0,0.15,1000)

rabi_S_sim = Rabi(
	system = sys,
	pulse_duration = tp_S,
	h1 = h1,
	pulse_params = {'f_pulse':w0_mw},
	pulse_shape = square_pulse
	)

rabi_S_sim.run()
\end{python}
with numpy imported as \texttt{np}.
The \texttt{run} method is used to execute the simulation, by setting and solving the coupled ODEs as described in Sec.~\ref{sec:dynamics}, where the results of the simulations are stored in the \texttt{results} attribute of the \texttt{rabi\_S\_sim} object. 
Here, the frequency of the excitation field $\bold{B}_1(t)$ is passed to the object through the \texttt{f\_pulse} key of the \texttt{pulse\_params} dictionary.

The resulting expectation value of the fluorescence observable (Eq.~\ref{eq:F}) as a function of the pulse length is shown in Fig.~\ref{fig:rabi}~(a), which can be plotted with the \texttt{plot\_results} method from the \texttt{Analysis} class.
In accordance with Ref.~\cite{polarization_population_2}, the electron spin performs full Rabi oscillations between the $m_S=0$ and $m_S=-1$ levels, conditioned to the $^{13}$C spin being at the $m_{I^c}=+1/2$ state.
The $m_S=+1$ levels and the nuclear states are not affected by this pulse, as expected.

Apart from a conditional rotation of the electron spin, the carbon nuclear spin can also be excited by setting \texttt{f\_pulse} in resonance with \texttt{w0\_rf}.
Now, we consider $\ket{m_S=-1}\otimes\ket{m_{I^c}=+1/2}$ as the initial state and define an observable for the nuclear spin as
\begin{equation}\label{eq:Fc_1}
 \hat{F}_{I^c} = \hat{\mathds{1}} \otimes \ket{+1/2}\bra{+1/2} .
\end{equation}
Experimentally, this is indirectly measured through the electron spin by electron-nuclear double magnetic resonance (ENDOR)~\cite{ENDOR}.
To model the decoherence, we assume $\hat{C}=\hat{I}^c_z$ as the collapse operator in Eq.~\ref{eq:lindblad}, with rate $\Gamma_2 = 0.5$~MHz.
Altogether, the time evolution of the nuclear spin within these considerations is simulated by:
\begin{python}
# Rabi nuclear simulation
sys.rho0 = tensor(basis(3,2),basis(2,0))
sys.observable = tensor(qeye(3),basis(2,0)*basis(2,0).dag())

gamma2 = .5
sys.c_ops = gamma2*tensor(qeye(3),jmat(1/2,'z'))

tp_I = np.linspace(0,2.5,1000)

rabi_I_sim = Rabi(
	system = sys,
	pulse_duration = tp_I,
	h1 = h1,
	pulse_params = {'f_pulse':w0_rf}
	)

rabi_I_sim.run()
\end{python}
where the \texttt{c\_ops} attribute of the \texttt{NV} instance, which was previously empty, is now set to the operator.
The \texttt{pulse\_shape} parameter is set to square pulses by default, which will be omitted from now on.

The results of the $\langle \hat{F}_{I^c} \rangle$ expectation value are shown in Fig.~\ref{fig:rabi}~(b).
Differing from the electron spin, the oscillation is damped by decoherence of the nuclear spin and has a much smaller Rabi frequency due to the smaller gyromagnetic ratio and thus, weaker coupling to the excitation field $\bold{B}_1(t)$.
The digital twin accurately reproduces the results presented in Ref.~\cite{polarization_population_2}, apart from the fact that the simulations show a much larger contrast.
This can be attributed to an imperfect initialization of the spins, the fundamentally limited NV readout mechanism and other experimental parameters related to the photon detection which are not accounted for in the simulation (Sec.~\ref{sec:optical}), but can be post-processed with the \texttt{ExpData} class.
Together with unconditional excitations, these two rotations form the computational basis for arbitrary quantum gates with the NV-$^{13}$C pair, which will be further explored in the subsequent sections.


\subsection{Sensing and Control of External Spins by Dynamical Decoupling}\label{sec:sensing}

\begin{figure*}[t!]
	\includegraphics[width=\textwidth]{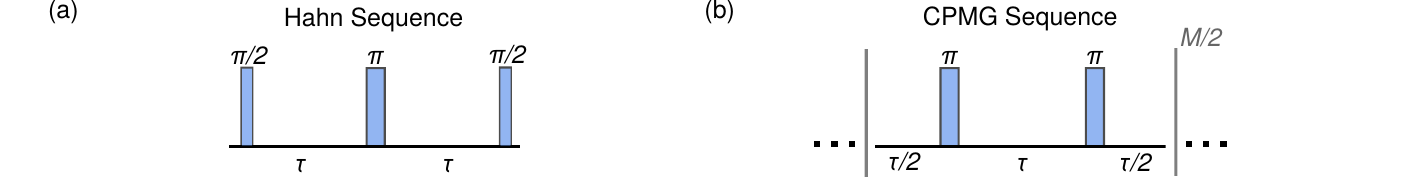}
	\includegraphics[width=\textwidth]{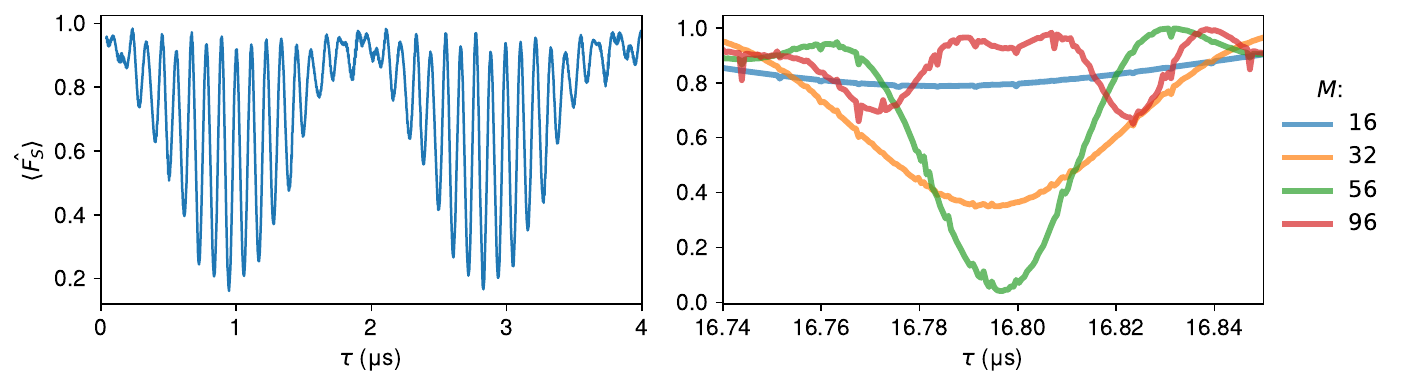}
	\caption{\textbf{(a)} Hahn echo sequence simulation of NV-$^{13}$C system as performed experimentally by L. Childress \textit{et al.} (2006)~\cite{RWA_2}.
	The fluorescence shows a characteristic spin echo envelope modulation, with a fast and a slow frequencies corresponding to the $^{13}$C nuclear spin Larmor frequencies at $\ket{m_S =+1}$ and $\ket{m_S =0}$ levels, respectively.
	The Hahn echo sequence refocuses static dephasings from the $^{14}$N nuclear spin and environment, enabling the detection of the $^{13}$C nuclear spin.
	\textbf{(b)} CPMG sequence simulation of a weakly coupled NV-$^{13}$C~\cite{13C_sensing_1} as experimentally performed by T. H. Taminiau \textit{et al.} (2012)~\cite{13C_sensing_1}.
	By repeating the $\pi$-pulse and free evolution of $\tau$ several times, the CPMG sequence is able to cancel out oscillating noises from the environment, enabling the sensing of the weakly coupled $^{13}$C nuclear spin.
	The intensities of the resonances show oscillations with the number of pulses $M$, which can be used to perform precise conditional gates to the $^{13}$C, exclusively by fast and precise MW pulses to the electron spin.
	}
	\label{fig:sensing}
\end{figure*}

The precise detection of weak spins in a noisy environment is a central challenge for quantum sensing.
At the same time, the coherent control of such magnetic momenta via a central spin is of great interest for quantum simulation~\cite{qsim_NV} and computing.
Although they have different goals, both applications can be achieved by dynamical decoupling of the spins from their noisy environment~\cite{ambiguous_resonances, 13C_sensing_1, DD_NV}.
In the end, the problem is then reduced to precisely modeling the interaction of the central spin (NV's electron) and the external spin ($^{13}$C nuclei) under multipulse dynamical decoupling sequences.
Following the discussion of the previous section, we first simulate the coherent dynamics of the NV-$^{13}$C system in a Hahn echo sequence, as experimentally studied by L. Childress \textit{et al.} (2006)~\cite{RWA_2}.
Subsequently, we simulate the detection and control of a weakly coupled $^{13}$C by a CPMG sequence, as performed by T. H. Taminiau \textit{et al.} (2012)~\cite{13C_sensing_1}.
Finally, we simulate the sensing of a classical field RF $\bold{B}_2(t)$ with the XY8 and RXY8 sequences, as shown by L. Tsunaki \textit{et al.} (2025)~\cite{ambiguous_resonances}.

Once more, we begin the simulation by instantiating the \texttt{NV} class with the experimental parameters as given in Ref.~\cite{RWA_2}:
\begin{python}
# System definition
sys1 = NV(
	B0 = 4.2,
	units_B0 = 'mT',
	theta = -45,
	units_angles = 'deg',
	N = 14,
	temp = 300,
	units_temp = 'K'
	)
\end{python}
where we use the \texttt{theta} argument to represent the misalignment of the magnetic field $\bold{B}_0$ with the NV axis $z$, which by default is 0.
Here, the magnetic field has a small amplitude of 4.2~mT along the $\bold{z}-\bold{x}$ direction.
In contrast to the application in Sec.~\ref{sec:rotations}, the nitrogen nuclear spin cannot be neglected, as it slightly shifts the energy levels of the electron spin, thus changing its Larmor frequency and affecting the envelope frequencies of the Hahn echo signal.
For that, we set the parameter \texttt{N} to 14 representing the $^{14}$N isotope.
If the system was otherwise composed with a $^{15}$N, additional signals from the nitrogen precession would be observed~\cite{ambiguous_resonances}.
However, this is not the case for $^{14}$N, due to the quadrupole interaction that fixes the nuclear spin orientation, thus suppressing its precession.
Lastly, for illustration purposes, we set the temperature to 300~K with the \texttt{temp} and \texttt{units\_temp} attributes, which calculates the initial state $\hat{\rho}_0$ from the Boltzmann distribution, as in Eq.~\ref{eq:BD_rho0} but for the $^{14}$N isotope.
If no input is given for \texttt{temp}, as in the other applications, the initial density matrix of the nuclear spin is assumed to be the identity matrix (Sec.~\ref{sec:optical}), where for these values of temperature and magnetic field the thermal polarization is minimum.

The coupling Hamiltonian, as in Eq.~\ref{eq:H2_quantum}, has a hyperfine coupling tensor in units of MHz given by
\begin{equation*}
	\bold{A}^c_{hf} =
	\begin{bmatrix*}[r]
		5.0 & -6.3 & -2.9 \\
		-6.3 & 4.2 & -2.3 \\
		-2.9 & -2.3 & 8.2
	\end{bmatrix*} ,
\end{equation*}
representing a weaker coupling compared to Sec.~\ref{sec:rotations}, but still prominent.
The representation of $\hat{H}_2$ and composition with the NV is the same as performed in the previous section.
Lastly, the Rabi frequency \texttt{w1} and the the frequency of the MW pulse excitations \texttt{w0} are defined as follows~\cite{RWA_2}:
\begin{python}
# Frequencies definition
w1 = 15
w0 = np.mean(sys1.energy_levels[12:]) - np.mean(sys1.energy_levels[1:6])
\end{python}
Here, the frequency of the MW pulse is calculated by subtracting the average of all six nuclear sublevels at $m_S=+1$ from the ones in $m_S=0$.
This way, with a short hard MW pulse, the $\ket{m_S=0} \leftrightarrow \ket{m_S=+1}$ transition is obtained, unconditional to the nuclear states, as opposed to Sec.~\ref{sec:rotations}.
This leads to a $\pi$-pulse duration shorter than $1/(2\omega_1)$, resulting from a hyperfine enhancement effect~\cite{phd_neumann}.
The simulated Rabi oscillation can be seen at the QuaCCAToo tutorials. 

Given the defined system with the experimental parameters from Ref.~\cite{RWA_2}, the simplest dynamical decoupling sequence for suppressing the noise from the environment is the Hahn echo technique, as schematically represented in Fig.~\ref{fig:sensing}~(a).
The sequence is composed by two free evolutions periods of duration $\tau$, with an intermediate $\pi$-pulse to the electron spin.
In addition, $\pi/2$-pulses are included before and after the sequence to drive the electron spin from and to the quantization axis.
In NMR, the last projective pulse is not required though, as the measurement is performed perpendicular to the quantization axis.
The intermediate $\pi$-pulse causes a refocusing of static or slow dephasing processes, such as those arising from the $^{14}$N nuclear spin.
This allows the coupled $^{13}$C nuclear spin to be probed by the NV.

For performing a Hahn echo simulation in the software, we use the \texttt{Hahn} class:
\begin{python}
# Hahn simulation
from quaccatoo import Hahn
	
tau_hahn = np.linspace(0.04,4,2000)
	
hahn_sim = Hahn(
	free_duration = tau_hahn,
	pi_pulse_duration = 0.0316,
	system = sys1,
	h1 = w1*sys1.MW_h1, 
	pulse_params = {'f_pulse':w0},
	projection_pulse = True,
	time_steps = 1000
	)

hahn_sim.run(map_kw={'num_cpus':32})
\end{python}
with the same arguments as defined previously.
In addition, \texttt{free\_duration} is the array containing the $\tau$ values to be simulated, \texttt{MW\_h1} is the attribute of the \texttt{sys1} object containing the standard MW control Hamiltonian operator $\sqrt{2} \hat{S_x}$  (Eq.~\ref{eq:h1}), \texttt{pi\_pulse\_duration} is the duration of the simulated $\pi$-pulse, \texttt{time\_steps} is the number of discrete time steps used in the simulation in each pulse, and \texttt{projection\_pulse} is a Boolean indicating to include a final project $\pi/2$-pulse or not, which is set to \texttt{True} by default and will be omitted from now on.
It is important to note, however, that the actual pulse separation is the $\tau$ variable with the duration of the $\pi$-pulse subtracted.
In most studies, perfect $\delta$-pulses are considered and both quantities coincide.
Finally, the simulation is executed with the \texttt{run} method, which takes a dictionary containing the options for the parallelization routines.
In this case, we include the \texttt{num\_cpus} key as an illustration, which limits the use of the hardware cores for the calculation, in order to avoid over-threading.
If the key in the dictionary parameter is not given, the parallelization will run on all of hardware cores by default.

The results of the Hahn echo simulation are shown in Fig.~\ref{fig:sensing}~(a), where we observe the characteristic  Electron Spin Echo Envelope Modulation (ESEEM)~\cite{ESEEM, RWA_2}, where the fluorescence observable of the electron spin shows two  frequencies.
The fast one corresponds to the Larmor frequency of the $^{13}$C nuclear spin at the $m_S=+1$ level, while the slower frequency corresponds to the one at $m_S=0$, which is simply given by the gyromagnetic ratio of the nuclear spin $\gamma^c$ due to the hyperfine coupling being negligible at $m_S=0$.
By measuring these modulation frequencies at different magnetic field configurations $\bold{B}_0$ , the hyperfine interaction matrix can be experimentally reconstructed for arbitrary spins hyperfinely coupled to a central spin~\cite{ambiguous_resonances}.

The Hahn echo sequence provides a simple mechanism for selective addressing and sensing spins from the environment via the NV center, as demonstrated here.
Nonetheless, the noise filtering can be significantly improved by repeating the $\pi$-pulse and free evolutions several times.
This is known as the CPMG sequence~\cite{CP_sequence, CPMG}, as shown in Fig.~\ref{fig:sensing}~(b).
The periodic reversals of the system's time evolution cancels out the oscillating dephasings and environmental noises from the total evolution, except for frequencies corresponding to twice the pulse separation $f_0=1/(2\tau)$ and odd multiples thereof \cite{multipulse1, spurious}.
As a result, weak oscillating magnetic fields from single nuclear spins can be filtered from a much stronger background, even at room temperature.
This represent a great improvement compared to superconducting quantum interference device (SQUID)~\cite{SQUID} sensing.

For simulating a CPMG sequence of an NV weakly coupled to a lattice $^{13}$C, we instantiate the object \texttt{sys2} with parameters $B_0=40.1$~mT, $\theta = 0$ and $N=14$ as in Ref.~\cite{13C_sensing_1}.
Furthermore the MW $\pi$-pulse duration is set to 10~ns, its frequency \texttt{w0} is in resonance with the $\ket{m_S=0} \leftrightarrow \ket{m_S=-1}$ transition.
The hyperfine coupling tensor is
\begin{equation*}
	\bold{A}^c_{hf} =
	\begin{bmatrix}
		0 & 0 & 0.044 \\
		0 & 0 & 0 \\
		0.044 & 0 & 0.032
	\end{bmatrix} ,
\end{equation*}
(in units of MHz), representing a much weaker coupling than in the previous case.
To run the CPMG simulation with \texttt{M} pulses, we do as follows:
\begin{python}
# CPMG simulation
sol_opt = {'atol':1e-16, 'rtol':1e-16, 'nsteps':1e8, 'order':30}
	
tau_cpmg = np.linspace(16.74,16.85,200)
	
cpmg_sim = CPMG(
	free_duration = tau_cpmg,
	pi_pulse_duration = 0.01,
	system = sys2,
	h1 = w1*sys2.MW_h1, 
	pulse_params = {'f_pulse':w0},
	M = M,
	time_steps = 1000,
	options = sol_opt
	)
	
cpmg_sim.run()
\end{python}
The \texttt{sol\_opt} variable contains a series of parameters for the ODE solver, which can be passed to the object through the \texttt{options} argument.
In this case, the absolute and relative tolerances are decreased, while the number of steps and order are increased, to have a more robust solution for such a long pulse sequence.

The simulation results of the CPMG sequence applied to the NV-$^{13}$C system are shown in Fig.~\ref{fig:sensing}~(b).
The fluorescence observable shows the 8\textsuperscript{th} order resonance from the weakly coupled $^{13}$C spin, displaying oscillations in its intensity as a function of the number of pulses $M$, due to changes in the filtering function of the multipulse sequence~\cite{phd_muller}.
By keeping a fixed pulse separation value $\tau$ and changing the number of pulses $M$, conditional gates can be applied to the nuclear spin exclusively through faster and more precise MW pulses to the electron spin~\cite{13C_sensing_1}.
This leads to an effective decoupling of the NV spin from arbitrary oscillating environmental noises, allowing a precise control of the weakly coupled $^{13}$C spin.
The simulation also presents a small numerical noise, resulting from truncation errors in the ODE solver at such long $\tau$ values.
Both numerical simulations of the Hahn echo and CPMG sequences match the experimental results from Refs.~\cite{RWA_2, 13C_sensing_1}, going beyond the semi-analytical model employed there, as it can be used for arbitrary coupling regimes, while considering realistic pulses.

\begin{figure}[t!]
	\includegraphics[width=\columnwidth]{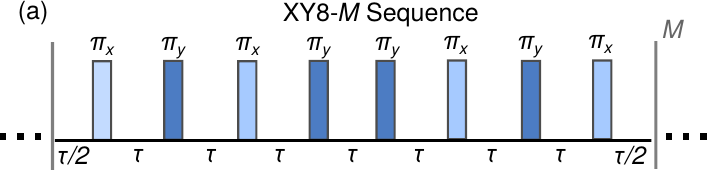}
	\includegraphics[width=\columnwidth]{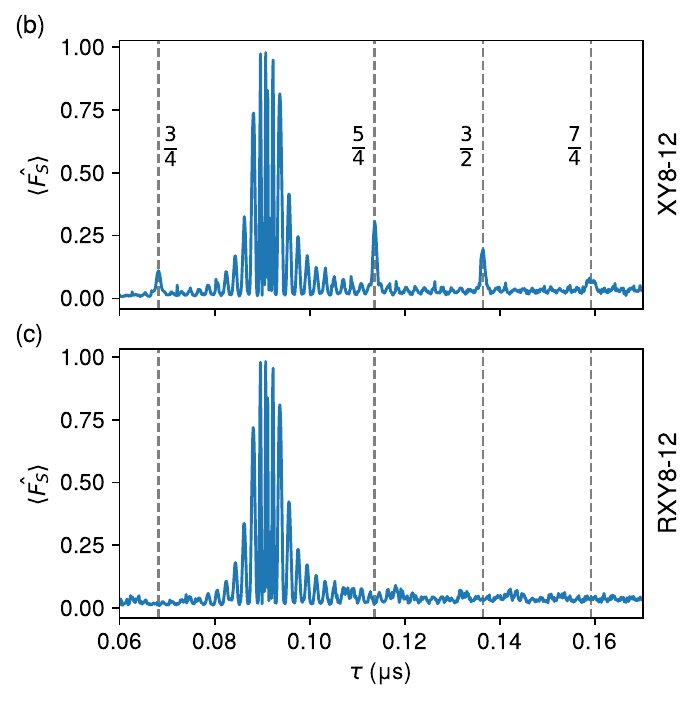}
	\caption{\textbf{(a)} Schematic of the XY8-12 pulse  sequence.
	It is composed of 8 intercalated $\pi$-pulses on $x$ and $y$ axes repeated $M$ times, thus canceling dephasings more efficiently than the CPMG sequence.
	\textbf{(b)} XY8-12 simulation of an NV sensing and external field $\bold{B}_2(t)$ with frequency $\omega_2=5.5$~MHz~\cite{ambiguous_resonances}.
	The NV's fluorescence observable shows a prominent resonant at $\tau_0=(2\omega_2)$ with multiple fringes.
	Apart from that, spurious harmonics are also present at fractions of $\tau_0$ due to the system's free evolution at the finite pulse lengths.
	\textbf{(c)} Simulation of RXY8-12 with a phase randomization in each XY8 block.
	With this random phase, the spurious harmonics are suppressed.
	This demonstrates the ability of the software to operate in arbitrary coupling regimes and time-dependent Hamiltonians.
	These simulations reproduce the experimental results presented by L. Tsunaki \textit{et al.} (2025)~\cite{ambiguous_resonances}.
	}
	\label{fig:xy8}	
\end{figure}

Finally, the CPMG sequence can be further improved to decouple noises acting on different axes by intercalating $\pi$-pulses over the $x$ and $y$ axes.
One of these sequences is the XY8-$M$~\cite{XY8}, with 8 anti-symmetrized $\pi_{x,y}$-pulses repeated $M$ times as schematically represented in Fig.~\ref{fig:xy8}~(a).
This way, the sequence has $8\times M$ pulses and can be used for more efficient sensing and control of external spins compared to CPMG.
In this case however, we consider the semi-classical interaction picture from Eq.~\ref{eq:H2_classical}, where a weak classical oscillating field $\bold{B}_2(t)$ is measured by the NV.
For that we instantiate \texttt{qsys3} with \texttt{N=15} and \texttt{B0=40} in mT.
We then define the sensing field as in Ref.~\cite{ambiguous_resonances} with $\gamma^e B_2 = 0.3$~MHz and $\omega_2=5.5$~MHz using:
\begin{python}
# B2(t) field definition
def B2(t, gamma_B2=5.5, w2=0.3):
	return gamma_B2*np.sin(w2*t)

H2 = [tensor(jmat(1,'z'), qeye(2)), B2]  
\end{python}
where \texttt{H2} follows the list-based definition of a time-dependent Hamiltonian from QuTip.
For performing the simulation, the \texttt{XY8} class from QuaCCAToo is instantiated with the experimental parameters and run:
\begin{python}
# XY8 simulation
from quaccatoo import XY8	
	
w1 = 20
tau_xy8 = np.linspace(0.06,0.17,1000)

xy8_sim  = XY8(
	M=12,
	free_duration = tau_xy8, 
	pi_pulse_duration = 1/2/w1,
	system = qsys3,
	h1 = w1*qsys3.MW_h1,
	pulse_params = {'f_pulse': qsys3.MW_freqs[0]},
	H2 = H2
	)

xy8_sim.run()
\end{python}
where the \texttt{H2} parameter (which was by default empty) is set with the time-dependent definition of $\hat{H}_2(t)$.
The frequency of the pulse \texttt{f\_pulse} is obtained from the \texttt{MW\_freqs[0]} attribute of the \texttt{NV} class, which calculates the frequency of the $\ket{m_S=0} \leftrightarrow \ket{m_S=-1}$ transition for the NV center of the selected nitrogen isotope, considering a hard unconditional pulse.

The simulation results of the $\bold{B}_2(t)$ sensing by XY8-12 are shown in Fig.~\ref{fig:xy8}~(b).
A prominent resonance with multiple fringes, resulting from the convolution of the filter function of the multipulse sequence with the spectral density of $\bold{B}_2(t)$~\cite{phd_muller}, is observed at $\tau_0=1/(2\omega_2)\cong 0.09$~\textmu{s}.
Apart from this resonance, the XY8 spectrum presents spurious harmonics at $3/4$, $5/4$, $3/2$ and $7/4$ fractions of the fundamental resonance at $\tau_0$.
These spurious harmonics originate from the free evolution of the system during the finite length of the pulses~\cite{spurious}, which can be an issue for quantum sensing applications, as they can be confused with other signals.
However, to suppress these, a random phase can be added to each XY8 block of the sequence~\cite{RXY8_1, RXY8_2}, known as the RXY8-$M$ sequence.

To simulate the RXY8-$12$ sensing of the external field $\bold{B}_2(t)$, the \texttt{RXY8} parameter is set to \texttt{True} in the \texttt{XY8} class, with the other parameters being the same as before.
The results of these simulations are shown in Fig.~\ref{fig:xy8}~(c).
In this case, the fundamental resonance is the same as for the XY8-12 sequence, but the spurious harmonics are suppressed, in accordance with Ref.~\cite{ambiguous_resonances}.
Altogether, this shows that our simulation software can operate with arbitrary coupling regimes (weak or strong, semi-classical or quantum) and with realistic imperfect pulses, taking into consideration the specific time-dependency of the $\hat{H}_1(t)$ and $\hat{H}_2(t)$ Hamiltonians.


\subsection{Quantum Teleportation}\label{sec:network}

While the first example in Sec.~\ref{sec:rotations} focuses on basic operations with the NV digital twin and Sec.~\ref{sec:sensing} discusses three applications related to sensing and computing, this section introduces a problem relevant to quantum communication and cryptography.
The process of quantum teleportation consists of transferring the state of one qubit to another spatially separated one, utilizing the non-local properties of quantum entanglement~\cite{QIP_NMR}.
This technique is an essential building block for quantum networks~\cite{NV_network_2, NV_network}.
Here, we simulate an adapted version of an unconditional teleportation between two NV centers, as originally demonstrated by W. Pfaff \textit{et al.} (2014)~\cite{NV_teleportation_1}.
Specifically, we focus on simulating a simplified version of the MW and RF pulses implementation, rather than the photonic component of the experiment, which by itself the greatest technical challenge of the experiment and the main source of errors.

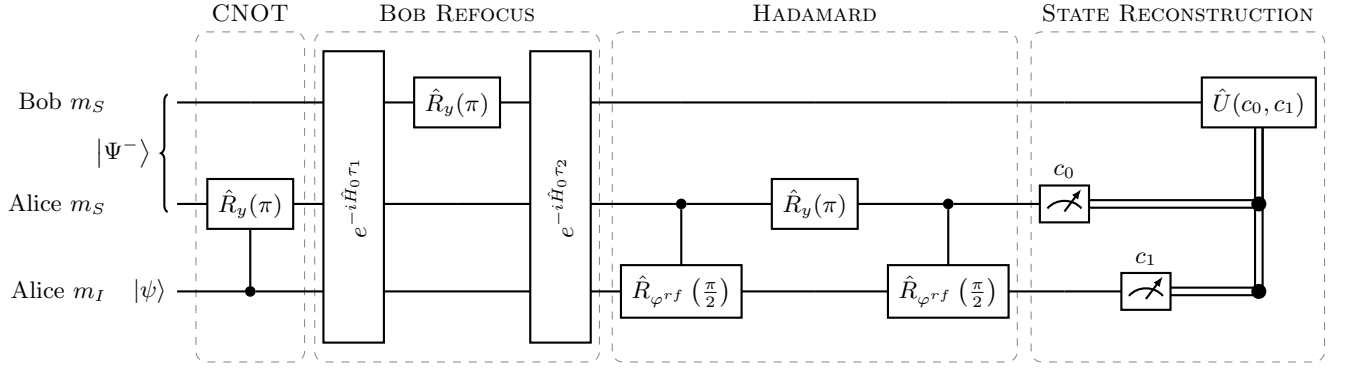
\begin{figure*}[t!]
	\begin{quantikz}[row sep=0.15cm, column sep=0.4cm]
		\lstick{Bob $m_S$} & &  \lstick[wires=2]{$\ket{\Psi^-}$} & \qw
		\gategroup[3,steps=1,style={dashed, color=gray!90, thin, rounded corners, inner xsep=1pt}, background]{{\sc CNOT}}
		& \gate[wires=3][.8cm]{\rotatebox{90}{$e^{-i \hat{H}_0 \tau_1}$}}\gategroup[3,steps=3,style={dashed, color=gray!90, thin, rounded corners, inner xsep=0pt}, background]{{\sc Bob Refocus}}
		& \gate{\hat{R}_y(\pi)}
		& \gate[wires=3][.8cm]{\rotatebox{90}{$e^{-i \hat{H}_0 \tau_2}$}}  & \qw
		\gategroup[3,steps=3,style={dashed, color=gray!90, thin, rounded corners, inner xsep=0pt}, background]{{\sc Hadamard}}
		& \qw & \qw & \qw
		\gategroup[3,steps=3,style={dashed, color=gray!90, thin, rounded corners, inner xsep=0pt}, background]{{\sc State Reconstruction}}
		& \qw & \gate{\hat{U}(c_0, c_1)}\\
		\lstick{Alice $m_S$} & & & \gate{\hat{R}_y(\pi)} & & \qw & & \ctrl{1} & \gate{\hat{R}_y(\pi)} &  \ctrl{1} & \meter{$c_0$} & \cw & \cwbend{-1} \\
		\lstick{Alice $m_I$} & & \lstick[wires=1]{$\ket{\psi}$} & \ctrl{-1} & & \qw & & \gate{\hat{R}_{\varphi^{rf}}\left(\frac{\pi}{2}\right)} & \qw & \gate{\hat{R}_{\varphi^{rf}}\left(\frac{\pi}{2}\right)} & \qw & \meter{$c_1$} & \cwbend{-2} 
	\end{quantikz}
	\caption{Circuit diagram of a teleportation protocol adapted from W. Pfaff \textit{et al.} (2014)~\cite{NV_teleportation_1}.
		Alice and Bob NV electron spins are initially in a Bell state $\ket{\Psi^-}$, while Alice's nitrogen spin is in a state $\ket{\psi}$ to be teleported.
		First, a CNOT gate is applied to Alice's electron spin conditioned to her nuclear spin state.
		Second, Bob performs a phase refocus of his qubit with a $\hat{R}_y(\pi)$ pulse between two free evolutions periods $\tau_1$ and $\tau_2$.
		Third, Alice performs a Hadamard gate on her nuclear spin, composed of three pulses, where the rotation axis $\varphi^{rf}$ for the nuclear pulses is optimized to avoid phase accumulation.
		Fourth, Alice measures both of her qubits yielding two classical bits $c_0$ and $c_1$, which also projects Bob's qubit into a known state.
		Finally, the classical bits are communicated to Bob, who performs an operation $\hat{U}(c_0, c_1)$ on his qubit, thus concluding the teleportation operation.}
	\label{fig:teleport_protocol}
\end{figure*}

This simple form of teleportation consists of two network parties: `Alice' and `Bob'.
Alice has a $^{14}$NV center with two qubits, the nitrogen nuclear spin and the electron spin.
Bob has another distant NV, where the nuclear spin can be neglected.
This system can be defined in QuaCCAToo using:
\begin{python}
# System definition	
NVb = NV(B0=18, units_B0='mT', N=0)
NVa = NV(B0=25, units_B0='mT', N=14)

NVb.truncate(mS=1)
NVa.truncate(mS=1, mI=1)

from quaccatoo import compose_sys
sys = compose_sys(NVb, NVa)
\end{python}
where we assume some typical values for $B_0$.
To avoid performance issues of simulating a large Hilbert space with dimension $d=3^3=27$ from the three spin-1 subsystems, the \texttt{truncate} method is used to exclude the $\ket{m_S=+1}$ and $\ket{m_I=+1}$ levels.
This, as seen in the previous examples, does not compromise the modeling of the system, as long as the transition frequencies are well separated compared to the spectral width of the excitation. 
This way, the system is defined within the subspaces $m_{S,I}=0,-1$, with a total dimension of $d=2^3=8$ instead.
Lastly, the \texttt{compose\_sys} function is used to compose the two NV centers from Alice and Bob, which internally calculates the new attributes of the \texttt{sys} object.

Initially, the two NVs' electron spins are in a maximally entangled Bell state $\ket{\Psi^-} = (\ket{01} - \ket{10})/\sqrt{2}$, where $\ket{0} \equiv \ket{m_S=0}$ and $\ket{1} \equiv \ket{m_S=-1}$.
While this is presented as a postulate for the simulation applications, the experimental generation of the Bell state poses a great technological challenge, given the poor spectral stability of NVs and high relative background counts.
This is achieved through a photon emission mixing by a beam splitter and mutual photon detection, which thus heralds an entanglement event for the two NVs' electron spins~\cite{heralded_entang}.
The optical properties of the NVs optical emission can be nonetheless improved by micro-engineering of the diamond surface~\cite{delta_doping, lens_1, lens_2}.

Apart from the initial state of the electron spin, the state $\ket{\psi} = \alpha \ket{0} + \beta \ket{1}$ to be teleported between Alice and Bob is initially stored in Alice's $^{14}$N spin.
Experimentally, it can be initialized into the state $\ket{1} \equiv \ket{m_I = -1}$ by a projective measurement of the electron spin~\cite{single_shot_readout, 14N_init}. 
The nuclear spin can then be brought to the state $\ket{\psi}$ with the appropriate resonance pulses~\cite{NV_teleportation_1}, as discussed in Sec.~\ref{sec:rotations}.
In the simulations, the initial state of the three spins $\ket{\Psi^-}\otimes\ket{\psi}$ is hence represented by:
\begin{python}
# Initial state definition
Psi_ = tensor(basis(2, 0),basis(2,1)) - tensor(basis(2,1),basis(2,0))

psi = alpha*basis(2,0)+beta*basis(2,1)

sys.rho0 = tensor(Psi_,psi).unit()
\end{python}
for given values of $\alpha, \beta \in  \mathbb{C}$.
By taking $\alpha=\beta=1/\sqrt{2}$ we have the initial state $\ket{+X}$, with $\alpha=1/\sqrt{2}$ and $\beta=i/\sqrt{2}$ we have $\ket{+Y}$, and finally with $\alpha=1$ and $\beta=0$ we have $\ket{+Z}$.
As in Sec.~\ref{sec:rotations}, the initial state in this case is pure.

Given the initial state, the whole teleportation protocol is schematically represented in Fig.~\ref{fig:teleport_protocol}.
Since there is no standard predefined sequence in QuaCCAToo for this protocol, we can simulate it by instantiating an object from \texttt{PulsedSim} with \texttt{sys} as an argument:
\begin{python}
# Sequence instantiation
from quaccatoo import PulsedSim
seq = PulsedSim(sys)
\end{python}
The first part of the teleportation protocol is a CNOT gate on Alice's electron spin conditioned to her nuclear spin.
This is expressed by a soft selective $\hat{R}_y(\pi)$ pulse of the $\ket{m_S=0} \leftrightarrow \ket{m_S=-1}$ electron transition for the $\ket{m_I=-1}$ nuclear state.
The intensity of this pulse needs to be low, in order to ensure a narrow bandwidth excitation.
Furthermore, the $\ket{m_I=0}$ states would ideally remain unchanged during the operation, but due to the system's evolution under the time-independent Hamiltonian $\hat{H}_0$ (Eq.~\ref{eq:H0}) during the finite pulse length, these states also accumulate a phase.
This phase is minimized by taking the Rabi frequency of the CNOT pulse as $\omega_1^{cnot} = a_\parallel^n/\sqrt{3}$~\cite{NV_teleportation_1}, where $a_\parallel^n$ is the parallel component of the hyperfine interaction as defined in Sec.~\ref{sec:H0}.
With these considerations, we can define the ODE solver options dictionary \texttt{sol\_opt}, the excitation frequency of the pulse \texttt{w0\_cnot}, the Rabi frequency \texttt{w1\_cnot}, the duration of the pulse \texttt{tpi\_cnot} and the $\hat{h}_1$ Hamiltonian for the CNOT (Eq.~\ref{eq:h1}) with:
\begin{python}
# CNOT gate parameters
sol_opt = {'nsteps':1e9}
w0_cnot = NVa.energy_levels[2]
w1_cnot = 2.14/3**.5
tpi_cnot = 1/(2*w1_cnot)
h1_cnot = w1_cnot*tensor(qeye(2),NVa.MW_h1)
\end{python}
where the transition frequency $\omega_0^{cnot}$ is obtained from the \texttt{energy\_levels} attribute of \texttt{NVa}.
A pulse with these parameters can be added to the sequences with the \texttt{add\_pulse} method using:
\begin{python}
# CNOT execution
seq.add_pulse(
	duration = tpi_cnot,
	h1 = h1_cnot,
	pulse_params = {'f_pulse':w0_cnot,
		'phi_t':np.pi/2},
	options = sol_opt
	)
\end{python}
By default, \texttt{add\_pulse} assumes a square pulse and 100 time steps in the time array discretization, which should not be confused with the \texttt{nsteps} key for the solver options dictionary that is increased to \texttt{1e9}.
In the lab frame, the $\phi$ angle of the rotation axis (Sec.~\ref{sec:H1}) is expressed in terms of the temporal phase of the excitation field \texttt{phi\_t}.
With this implementation, the simulation is already executed upon calling \texttt{add\_pulse}, differently from predefined methods which require the \texttt{run} command.
However, if the user desires to execute a sequence in parallel whereas changing a certain variable, it can be defined within a function and called with \texttt{run}.

One of the main challenges of the pulse implementation of the protocol is the loss of the relative phase between the qubits due to the dynamics from $\hat{H}_0$.
Specifically, all three spins precess with different Larmor frequency $\omega_0$ during the finite pulse lengths, making it necessary to correct for the phase accumulation in Bob's qubit while Alice performs the teleportation operation on her qubits.
For that, we implement a refocusing scheme similar to a Hahn echo (Sec.~\ref{sec:sensing}), with a free evolution of $\tau_1$ followed by a $\hat{R}_y(\pi)$ pulse and another free evolution of $\tau_2$.
The free evolution times $\tau_1$ and $\tau_2$ are chosen such that the $\hat{R}_y(\pi)$ is exactly in the middle of the whole protocol duration prior to the state reconstruction, and thus Bob's qubit ends the protocol with the same state as he initially had.
This way, we have the first free evolution time given by 
\begin{equation*}
	\tau_1 + t_\pi^{cnot} + \frac{t_\pi^B}{2} = \frac{T}{2} ,
\end{equation*}
where $T$ is the total duration of the protocol.
To implement the refocusing scheme, we define the pulse parameters for Bob's qubit and use the \texttt{add\_free\_evolution} and \texttt{add\_pulse} methods:
\begin{python}
# Refocus scheme
w1_mwb = 22
tpi_mwb = 1/(2*w1_mwb)
w0_mwb = NVb.MW_freqs[0]
h1_mwb = w1_mwb*tensor(NVb.MW_h1,qeye(2),qeye(2))

seq.add_free_evolution(tau1)
seq.add_pulse(
	duration = tpi_mwb,
	h1 = h1_mwb,
	pulse_params = {'f_pulse': w0_mwb,
		'phi_t':np.pi/2},
	options = sol_opt
	)
seq.add_free_evolution(tau2)
\end{python}
where \texttt{tau1} and \texttt{tau2} are defined with their respective values, with $\tau_2$ as calculated below.
In Ref.~\cite{NV_teleportation_1}, the phase of Bob's qubit is protected with an XY4 sequence, as introduced in Sec.~\ref{sec:sensing}.
However, in this simulation, the implementation would become more complex than the straight experimental realization with multi-channel pulse generation, due to the intricate time placement of the $\pi$-pulses on Bob's qubit.
In the end, this difference in the refocusing scheme will require small adjustments to Bob's state reconstruction pulses, as discussed in the following analysis.

The next step of the protocol is a Hadamard gate on Alice's nuclear spin, which can be decomposed with two rotations, such as $\hat{R}_{y}(\pi/2)\hat{R}_{z}(\pi)$.
In this case, the first rotation on the $z$-axis responsible for correcting the phase can be omitted, hence configuring a pseudo-Hadamard gate.
In addition, due to the fact that the transition frequency of the nuclear spin depends on the $m_S$ state [Fig.~\ref{fig:energy_levels}~(b)], the gate can be composed by three pulses: a conditional rotation of $\hat{R}_{\varphi^{rf}}(\pi/2)$ on the nuclear spin over some chosen axis $\varphi^{rf}$, an unconditional rotation of $\hat{R}_y(\pi)$ on Alice's spin, and another rotation of $\hat{R}_{\varphi^{rf}}(\pi/2)$ on the nuclear spin~\cite{NV_teleportation_1}.
This way, the first pulse changes the nuclear state for $m_S=-1$, then the electron spin is inverted and the same RF pulses is applied again.
Assuming some typical values for the pulse parameters, where the Rabi frequency of Alice's electron spin for a hard pulse \texttt{w1\_mwa} is now much larger than for the CNOT gate, the three pulses are simulated with:
\begin{python}
# Hadamard gate
w0_rf = NVa.RF_freqs[2]
w0_mwa = NVa.MW_freqs[0]
w1_rf = .2
w1_mwa = 16
tpi_rf = 1/(2*w1_rf)
tpi_mwa = 1/(2*w1_mwa)
h1_rf = w1_rf*tensor(qeye(2),NVa.RF_h1)
h1_mwa = w1_mwa*tensor(qeye(2),NVa.MW_h1)

seq.add_pulse(
	duration = tpi_rf/2,
	h1 = h1_rf,
	pulse_params={'f_pulse':w0_rf,
		'phi_t':phi_rf},
	options=sol_opt
	)
seq.add_pulse(
	duration = tpi_mwa,
	h1 = h1_mwa,
	pulse_params={'f_pulse': w0_mwa,
		'phi_t':np.pi/2},
	options=sol_opt
	)
seq.add_pulse(
	duration = tpi_rf/2,
	h1 = h1_rf,
	pulse_params={'f_pulse':w0_rf,
		'phi_t':phi_rf},
	options=sol_opt
	)
\end{python}
The rotation axis \texttt{phi\_rf} ($\varphi^{rf}$) and the free evolution time $\tau_2$ need to be optimized for avoiding again that nuclear spin accumulates phases during the gate.
Thus, following the calibration procedure as discussed in detail in Ref.~\cite{NV_network} and shown in QuaCCAToo tutorials, we obtain $\varphi^{rf}=2.95$~radians ($169^\circ$).
By further optimizing the final teleportation fidelity, we choose the total refocusing procedure duration to be $\tau_1 + \tau_2 + t_\pi^{MW_a} = 2.26$~\textmu{s}, which then determines the value of $\tau_2$.
In contrast, dephasings of Alice's electron spin are already being refocused by the intermediate $\pi$-pulse, as in the Hahn sequence used in Sec.~\ref{sec:sensing}.

\begin{table}[t!]
	\begin{tabular}{|c||c|c|c|c|}
		\hline
		\multirow{2}{*}{$\ket{\psi}$} & \multicolumn{4}{c|}{Outputs $c_0 \, c_1$} \\ \cline{2-5}
		& 00 & 01 & 10 & 11 \\ \hline \hline
		\multirow{2}{*}{$\ket{+X}$} & $\hat{R}_{-y}\left(\frac{\pi}{2}\right)$ & $\hat{R}_{z}(\pi)\hat{R}_{y}\left(\frac{\pi}{2}\right)$ & $\hat{R}_{z}(\pi)\hat{R}_{y}\left(\frac{\pi}{2}\right)$ & $\hat{R}_{-y}\left(\frac{\pi}{2}\right)$ \\
		& 0.9624 & 0.9814 & 0.9889 & 0.9878 \\ \hline
		\multirow{2}{*}{$\ket{+Y}$} & $\hat{R}_{z}(\pi)\hat{R}_{x}\left(\frac{\pi}{2}\right)$ & $\hat{R}_{-x}\left(\frac{\pi}{2}\right)$ & $\hat{R}_{z}(\pi)\hat{R}_{x}\left(\frac{\pi}{2}\right)$ & $\hat{R}_{-x}\left(\frac{\pi}{2}\right)$ \\
		& 0.9585 & 0.9713 & 0.9779 & 0.9945 \\ \hline
		\multirow{2}{*}{$\ket{+Z}$} & $\hat{R}_{y}(\pi)$ & $\hat{R}_{y}(\pi)$ & $\hat{\mathds{1}}$ & $\hat{\mathds{1}}$ \\
		& 0.9999 & 0.9998 & 0.9985 & 0.9981 \\ \hline
	\end{tabular}
	\caption{Bob's state reconstruction operations $\hat{U}(c_0, c_1)$ and resulting fidelities for each input states and classical outputs measured by Alice.
	All inputs and outputs show fidelities close to 1.
	Comparatively to each other, the $\ket{+Z}$ state shows higher values, due to the fact that the state is less prone to dephasings in the plane perpendicular to the quantization axis.
	}
	\label{tab:U_fidelities}
\end{table}

At this point, the teleportation of the $\ket{\psi} = \alpha \ket{0} + \beta \ket{1}$ state to Bob's NV has already taken place, who has a superposition state composed of four different possible combinations with the coefficients $\alpha$ and $\beta$~\cite{QIP_NMR}.
Thus, with Alice measuring her qubits, Bob's qubit will be projected in one of the four combinations.
Physically, the electron spin is measured by the fluorescence observable, as in Eq.~\ref{eq:F_S}, while the nuclear spin is measured using the electron spin as an auxiliary qubit~\cite{single_shot_readout}, which gives rise to an observable as
\begin{equation*}
	\hat{F}_{I^n} = \hat{\mathds{1}} \otimes \hat{\mathds{1}} \otimes \ket{1}\bra{1} .
\end{equation*}
Given the two observables, Alice's measurements are implemented with the \texttt{measure\_qsys} method:
\begin{python}
# Alice measurements
obs0 = tensor(qeye(2),fock_dm(2, 0),qeye(2))
c0 = 1 - seq.measure_qsys(observable=obs0)

obs1 = tensor(qeye(2),qeye(2),fock_dm(2, 1))
c1 = seq.measure_qsys(observable=obs1)
\end{python}
with \texttt{fock\_dm} imported from QuTip.
Besides collapsing the system in one of the eigenstates of the observables, the two measurements performed with the \texttt{measure\_qsys} method yield two classical bits \texttt{c0} ($c_0$) and \texttt{c1} ($c_1$).
These two classical bits are then communicated to Bob via a classical channel, who now needs to reconstruct the $\ket{\psi}$ state on his qubit with some operation $\hat{U}(c_0, c_1)$, which also depends on the input state $\ket{\psi}$ in this application~\cite{NV_teleportation_1}.
The corresponding operations $\hat{U}(c_0, c_1)$ and their resulting fidelity for each possible $c_0$ and $c_1$ and input states $\ket{+X}$, $\ket{+Y}$ and $\ket{+Z}$ are shown in Tab.~\ref{tab:U_fidelities}, which slightly differ from Ref.~\cite{NV_teleportation_1} due to the different refocusing scheme and the qubit ordering.

\begin{figure*}[t!]
	\includegraphics[width=.32\textwidth]{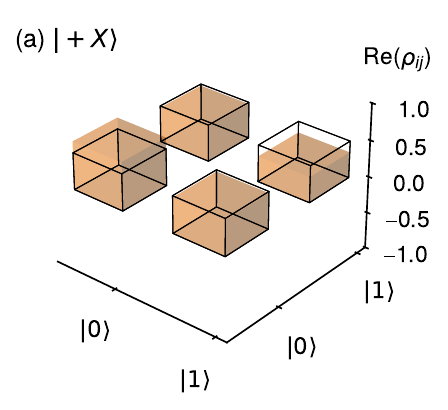}
	\includegraphics[width=.33\textwidth]{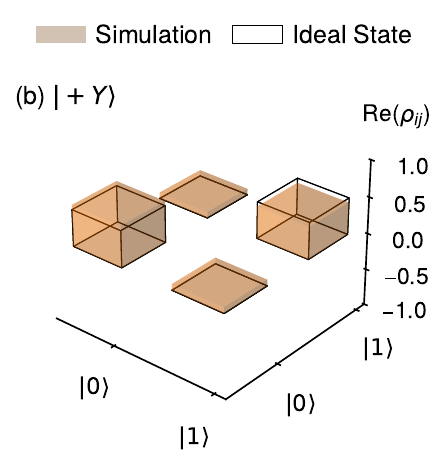}
	\includegraphics[width=.32\textwidth]{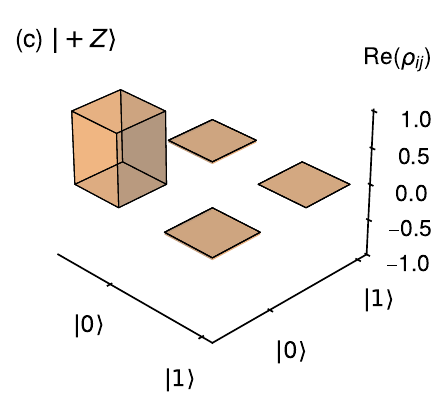}\vspace{-.8cm}
	\includegraphics[width=.32\textwidth]{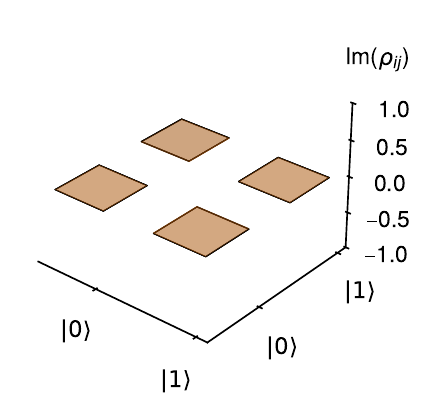}
	\includegraphics[width=.32\textwidth]{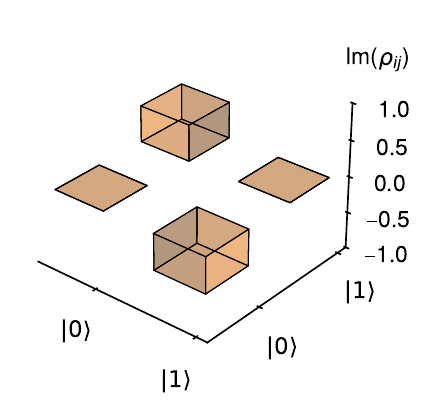}
	\includegraphics[width=.32\textwidth]{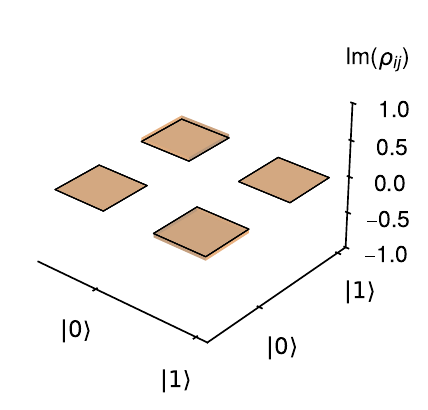}
	\caption{Simulated density matrices from teleportation protocol compared to the ideal teleported states of \textbf{(a)} $\ket{+X}$, \textbf{(b)} $\ket{+Y}$ and \textbf{(c)} $\ket{+Z}$.
	Overall, the simulations show a good agreement with the ideal state, with small deviations in the diagonal elements of the $\ket{+X}$ and $\ket{+Y}$ input states due to dephasings between the three qubits, with the resulting fidelities being shown in Tab.~\ref{tab:U_fidelities}.
	Such simulated fidelities are larger than the experimental values obtained by W. Pfaff \textit{et al.} (2014)~\cite{NV_teleportation_1}, in view of the fact that errors in the state initialization and readout are not considered.
	The principles involved in the teleportation protocol serve as the basis for quantum network applications.
	}
	\label{fig:teleport_results}
\end{figure*}

To exemplify the state reconstruction process, we take the $\ket{+Z}$ input state as an example.
If Alice's electron spin qubit is measured in the $\ket{1}$ state yielding $c_0, \, c_1 = 1, \, 0$ or $c_0, \, c_1 = 1, \, 1$, Bob does not need to perform any operation on his qubit, which is already in the state $\ket{\psi}$.
However, if Alice measures her electron spin in the state $\ket{0}$ with $c_0, \, c_1 = 0, \, 1$ or $c_0, \, c_1 = 0, \, 0$, then Bob needs to perform a rotation of $\hat{R}_y(\pi)$ so that his qubit is then in the state $\ket{\psi}$.
Altogether then, the state reconstruction for an initial state $\ket{+Z}$ is implemented with:
\begin{python}
# State reconstruction
if c0 == 0. and c1 == 0.:
	seq.add_pulse(tpi_mwb, h1_mwb, pulse_params={'f_pulse':w0_mwb, 'phi_t':np.pi/2}, options=sol_opt)
elif c0 == 0. and c1 == 1.:
	seq.add_pulse(tpi_mwb, h1_mwb, pulse_params={'f_pulse':w0_mwb, 'phi_t':np.pi/2}, options=sol_opt)
elif c0 == 1. and c1 == 0.:
	pass
elif c0 == 1. and c1 == 1.:
	pass
\end{python}
In the other two initial inputs however, $\pi/2$ pulses are also required to reconstruct the teleported state.
Where in these cases, the phase accumulated by Bob's qubit alters the final state and needs to be corrected, which is not the case for the $\pi$-pulse with a complete population inversion.
This can be achieved by adding a free evolution of duration $\tau_3$ such that the qubit will have completed an integer number of Larmor precessions and thus will have returned to the same state.
Mathematically, this means that the free evolution duration to cancel the phase accumulation must be
\begin{equation*}
	\tau_3 = \frac{\lceil \omega_0^B t_\pi^B/2 \rceil}{\omega_0^B} - \frac{t_\pi^B}{2} ,
\end{equation*}
where $\omega_0^B$ is the Larmor frequency for Bob's NV and $t_\pi^B$ the $\pi$-pulse duration, as previously defined. Here, $\lceil \cdot \rceil$ denotes the ceiling function, which rounds the value up to the nearest integer.
Finally, the $\pi$ rotations around the $z$-axis $\hat{R}_z(\pi)$ present in some $\hat{U}(c_0, c_1)$ can also be obtained simply with a free evolution of duration $1/(2\omega_0^B)$.

The simulated density matrices for each input state $\ket{\psi}$ and an arbitrary output $c_0c_1$ are visualized in Fig.~\ref{fig:teleport_results}, which can be plotted via QuaCCAToo with the \texttt{plot\_histogram} function.
The real and imaginary components are compared with the ideal teleported states of \textbf{(a)} $\ket{+X}$, \textbf{(b)} $\ket{+Y}$ and \textbf{(c)} $\ket{+Z}$, where we observe overall a strong agreement.
Small deviations can be seen in the diagonal elements of the $\ket{+X}$ and $\ket{+Y}$ states, which can be attributed to phase accumulations.
The corresponding fidelities for each inputs and outputs are given in Tab.~\ref{tab:U_fidelities}. Their values being larger than the corresponding experimental values provided in Ref.~\cite{NV_teleportation_1} can be attributed to the fact that this simulation does not consider errors in the optical initialization of the entangled state, neither the spin readout, nor its decoherence.
Nonetheless, this application complements the original work with a rigorous simulation of the MW and RF pulses components of the digital twin.

By avoiding the adoption of a rotating frame, a new layer of complexity to the dynamics of the system is introduced, related to the phase accumulations of each qubit in the laboratory frame.
These phase accumulations are often overlooked in perturbative models, such as the RWA, which may lead to discrepancies in experimental results~\cite{ambiguous_resonances, strong_driving_2, beyond_RWA}.
Another interesting result can be observed by comparing the fidelities for each input state.
The $\ket{+Z}$ state along the quantization axis of the NVs presents higher fidelities, due to the fact that this state is less prone to the phase accumulations in the perpendicular plane of the quantization.
\section{Conclusion and Outlook}\label{sec:conclusion}

Color centers are prominent physical platforms in the second quantum revolution, with groundbreaking applications within computing~\cite{QEC_NV,qubit_NV,optimal_control_1,13C_sensing_1,beyond_RWA}, sensing~\cite{RWA_2, ambiguous_resonances, NV_hamiltonian2, cancer1, cancer2, 13C_1, multipulse1}, networks~\cite{NV_teleportation_1, NV_network_2, NV_network} and memory-tokens~\cite{qtoken_1, qtoken_2}.
The interaction of the quantum-mechanical spin with optical fields can enable simple experimental control with a great potential for scalability~\cite{SiV_network} and integration with other systems.
At the same time, their solid state host can provide chemical and mechanical stability~\cite{biological_compatibility}, combined with long quantum state lifetimes~\cite{qubit_NV}.
Among the color center systems, NVs have been an epicenter of research, given their simple and robust optical initialization and readout mechanism~\cite{optical_pumping, optical_pumping_2}, in addition to the precise spin manipulation by MW and RF pulses~\cite{rabi, RWA_2, polarization_population_2}.
A free and open-source software for simulating the NV system and its interaction with the environment is therefore vital for the continued progress of the field.
Simplicity and usability are also essential for educational purposes.

In this work, we have rigorously modeled the NV Hamiltonian numerically with its environmental inputs, both its time-independent internal component $\hat{H}_0$, as well as its interaction with control fields $\hat{H}_1(t)$ and sensing targets $\hat{H}_2(t)$.
The description of the time evolution dynamics under MW and RF control fields makes no use of perturbative approximation methods, which can lead to discrepancies between theory and experiments~\cite{ambiguous_resonances}.
The simulation framework is based on the solution of the Lindblad equations within the static laboratory frame, which can account for non-unitary processes, such as relaxation and decoherence.

The optical inputs and outputs are not actually simulated in this work, but rather used as postulates for the states initialization and measurement observable.
While the exact simulation of the NV's interaction with the optical field can bring important insights into its operation~\cite{NV_optical_interaction, PES}, other systems heavily rely on the rigorous mathematical description of this interaction with light, such as SiV centers in diamond.
This way, in order for the simulation software to operate with arbitrary NV applications and other color centers, a proper quantum-mechanical description of the spins and Fock space interaction needs to be modeled.
Or within a semi-classical approach of the electric field interaction with an electric dipole moment, as used for spins in magnetic field.
Furthermore, the rest of this work can be readily implemented with other color centers through the use of QuaCCAToo, by adjusting the Hamiltonians, observables and initial states of the desired system.

The usability, robustness, range and flexibility of our framework are attested by the three application examples provided here.
Within the scope of quantum computing and sensing, Secs.~\ref{sec:rotations} and \ref{sec:sensing}, we have shown that the software precisely describes the NV's (hyperfine) interaction with externally coupled spins in different regimes.
This enables the simulation of long and complex dynamical decoupling sequences, as well as simpler conditional and unconditional excitations by the control field.
The use of realistic finite-length MW and RF pulses gives rise to interesting physical phenomena, which would otherwise not be accounted for in widely adopted approximations of $\delta$-pulses with no temporal length.
All of these simulations are implemented with QuaCCAToo's pre-built pulsed sequences and methods, with a concise and intuitive usability.

In Sec.~\ref{sec:network}, we expand the scope of the digital twin with a simulation of a quantum network application,
where a quantum state is teleported between two entangled NV spins, within a custom built QuaCCAToo pulsed sequence.
Once more, the solution of the system dynamics in the laboratory frame brings new physical elements to the simulation related to the relative phase accumulated by the qubits due to the effect of $\hat{H}_0$ during finite pules, which are mostly overlooked by the adoption of rotating frames.
All these applications also serve to validate our software, by showing robust correlations with the experimental results from previously established groundbreaking reports~\cite{polarization_population_2, RWA_2, 13C_sensing_1, ambiguous_resonances, NV_teleportation_1}.
Furthermore, this work complements the original experimental studies with a rigorous numerical modeling of the pulse implementation in the laboratory frame.
Where, as demonstrated here, in the ambiguous resonances of DD sequences in Sec.~\ref{sec:sensing} and the phase accumulation in the teleportation from Sec.~\ref{sec:network}, oversimplifications of the system's dynamics under the control fields can overlook important physical effects~\cite{ambiguous_resonances, spurious, strong_driving1, strong_driving_2}.

Apart from the absence of a complete simulation of the NV interaction with optical fields, other limitations of the digital twin are related to Python's limited performance compared to compiled programming languages, such as C++ and Fortran.
On the other hand, Python holds unmatched accessibility, provided by its simple and human-readable syntax.
An alternative high-level dynamic programming language with better performance than Python, but with similar accessibility, would be the fast evolving Julia language~\cite{julia}.
While this could potentially improve the software's efficiency, it requires testing and these mathematical problems are nonetheless limited by their intrinsic single threaded nature.
That is, for calculating the quantum state in a later time step, one needs to calculate it in all previous elements of the time array in sequence.
This inherently limits the use of parallelization procedures in the CPU only for different variables in the sequence (such as the frequency of the pulse or the pulse separation values) and not the time array itself.
Alternatively, performance improvements could be gained from new theoretical developments in quantum master equation solvers~\cite{QME}, such as Dyson series based solvers~\cite{dyson_series}.


\section*{Data and Code Availability}

All experimental data and simulations used in this work are available at the QuaCCAToo Github repository~\cite{quaccatoo}.
All simulations were done with version number 1.1.0 of the QuaCCAToo software and due to its active development, these implementation might become outdated.

\section*{Acknowledgments}

We acknowledge Dr. Kseniia Volkova from Helmholtz-Zentrum Berlin for designing the QuaCCAToo logo.
We are also grateful to Simon Sekavčnik from the Technical University of Munich for the insightful discussions on the concept of the digital twin.
This work was supported by the \textit{Bundesministerium für Bildung und Forschung} (BMBF) under the project \textit{DIamant-basiert QuantenTOKen} (DIQTOK - n\textsuperscript{o} 16KISQ034) and Diamant-basierte Quantenmaterialien (DIAQUAM - n\textsuperscript{o} 13N16956).
In addition, this work received funding from the \textit{Deutsche Forschungsgemeinschaft} (DFG) grant 410866378.


\bibliographystyle{apsrev4-2}

\end{document}